\documentclass[a4paper,onecolumn,11pt,accepted=2022-12-22]{quantumarticle}
\pdfoutput=1
\usepackage[utf8]{inputenc}
\usepackage[english]{babel}
\usepackage[T1]{fontenc}
\usepackage{amsmath, amssymb}
\usepackage{hyperref}

\usepackage{cite}
\usepackage{lipsum}

\usepackage{verbatim}
\usepackage{physics}
\usepackage[numbers]{natbib}

\begin{document}

\title{Solvable model of deep thermalization with distinct design times}

\author{Matteo Ippoliti}
\affiliation{Department of Physics, Stanford University, Stanford, CA 94305, USA}
\email{ippoliti@stanford.edu}
\orcid{0000-0003-4095-0038}
\author{Wen Wei Ho}
\affiliation{Department of Physics, Stanford University, Stanford, CA 94305, USA}
\affiliation{Department of Physics, National University of Singapore, Singapore 117542}
\email{wenweiho@nus.edu.sg}
\orcid{0000-0001-7702-2900}
\maketitle

\begin{abstract}
We study the emergence over time of a universal, uniform distribution of quantum states supported on a finite subsystem, induced by projectively measuring the rest of the system.
Dubbed {\it deep thermalization}, this phenomenon represents a form of equilibration in quantum many-body systems stronger than regular thermalization, which only constrains the ensemble-averaged values of observables.
While there exist quantum circuit models of dynamics in one dimension where this phenomenon can be shown to arise exactly, these  are special in that deep thermalization occurs at precisely the same time as regular thermalization. 
Here, we present an exactly-solvable model of chaotic dynamics where the two processes can be shown to occur over different time scales.
The model is composed of a finite subsystem coupled to an infinite  random-matrix bath through a small constriction, and highlights the role of locality and imperfect thermalization in constraining the formation of such universal wavefunction distributions.
We test our analytical predictions against exact numerical simulations, finding excellent agreement. 
\end{abstract}


\section{Introduction \label{sec:intro}}

The advent of quantum simulators---systems of many quantum particles whose interactions are finely programmable~\cite{altman_quantum_2021}---is beginning to allow the exploration of novel, complex many-body phenomena, particularly out of equilibrium~\cite{bernien_probing_2017, choi_observation_2017, mi_time-crystalline_2022, randall_many-bodylocalized_2021, dumitrescu_dynamical_2022}.
Beyond their advanced control capabilities, 
one unique aspect of quantum simulators is the nature of the information they provide on the underlying quantum state.
In many physical realizations (e.g., ultracold atoms with quantum gas microscopes, individually trapped ions or Rydberg atoms, superconducting qubits etc.), measurements involve the simultaneous read-out of the state of individual degrees of freedom, such as the occupation number of particles at a site in an optical lattice, or whether an atom is in its ground or excited state.
In such cases, we can think of a measurement in a given run of the experiment as obtaining a classical `snapshot' of the configuration of the entire system. 
Running an experiment multiple times, we then get a collection of snapshots, which constitutes {\it microscopically-resolved} information about the state. This represents a fundamentally new way of interrogating quantum many-body systems (as compared to more traditional macroscopic, coarse-grained probes, such as linear response to external perturbations or expectation values of order parameters).
The vast amount of information that is  generated by this type of probe immediately raises a number of questions.
For example, what universal phenomena can we uncover based on the information content of these bit-strings? Can such information be used to characterize novel phases of matter, whether in or out of equilibrium?
And what sorts of applications, especially for quantum information science, can we develop that leverage such large amounts of data?

There have been many recent developments along these fronts. 
Measuring local degrees of freedom during an otherwise-unitary many-body evolution was found to give rise to a novel kind of nonequilibrium, `measurement-induced' phases and associated phase transitions, driven by a competition between entanglement creation and destruction~\cite{skinner_measurement-induced_2019, li_quantum_2018, choi_quantum_2020, gullans_dynamical_2020, ippoliti_entanglement_2021, potter_entanglement_2021, fisher_random_2022, noel_measurement-induced_2022, koh_experimental_2022}.
Advanced data processing methods have been developed which allow for the efficient extraction of many properties of a quantum system by manipulating only compact, classical descriptions of its state, dubbed {\it `classical shadows'}~\cite{aaronson_shadow_2018, huang_predicting_2020, elben_randomized_2022}, which are constructed from snapshots in locally-randomized measurement bases. 
In the context of digital quantum simulators, the statistical analysis of classical snapshots is at the heart of the {\it random circuit sampling} problem, an ideal test of quantum computational advantage with noisy, intermediate-scale quantum hardware, which has been the subject of intense theoretical and experimental study in recent years~\cite{arute_quantum_2019, wu_strong_2021, aaronson_complexity-theoretic_2017, zhou_what_2020, gao_limitations_2021, deshpande_tight_2021, dalzell_random_2021}.
Relatedly, methods based on the statistics of such classical snapshots have been proposed to quantify the many-body fidelity between two quantum states, allowing the benchmarking of quantum simulator performance~\cite{choi_emergent_2021, mark_benchmarking_2022}. 

In this work, we consider a novel nonequilibrium universality enabled by this measurement paradigm, 
which can be understood as a generalization of the concept of 
quantum thermalization in isolated quantum many-body systems---the process by which local subsystems approach thermal equilibrium despite the global dynamics being unitary and thus reversible~\cite{srednicki_chaos_1994, rigol_thermalization_2008, abanin_colloquium_2019, alhambra_quantum_2022}.
Conventionally, such a phenomenon is explained by the fact that a local subsystem $A$ typically becomes entangled with its surroundings $B$ (the `bath', which is taken to be large), over time.
Ignoring the precise state of the bath, the state on $A$ is then described by a statistical mixture of pure quantum states which replicates thermal equilibrium predictions on average (i.e., expectation values in an appropriate Gibbs ensemble), in accordance with statistical mechanical principles. 
However, the measurement capabilities of quantum simulators, as described above, motivate a refinement of the above concept:
it is not necessary to assume that the bath is inaccessible; instead, one can study properties of the local subsystem $A$ {\it conditioned} upon  particular configurations of the bath $B$, information which is accessible from global measurement snapshots. 
In particular, one can imagine obtaining (classical) information of the state of $B$, which can then be used to index a {\it pure} quantum state on $A$ arising from the collapse of the global wavefunction after the measurement; the collection of all such quantum states is called the {\it projected ensemble}~\cite{cotler_emergent_2021, choi_emergent_2021}.

Much of the interest in studying these ensembles lies in the possibility that they might exhibit a universal limiting form at late times.
Indeed, it has been conjectured~\cite{cotler_emergent_2021, choi_emergent_2021} that for ensembles built from quantum many-body states undergoing chaotic quench dynamics, and in the absence of conservation laws or at very high energy density (corresponding to infinite temperature), the limiting form should be the unitarily-invariant (i.e., Haar) distribution over the Hilbert space---i.e., any pure state should be as likely as any other. 
One can view this as a generalization of the fundamental statistical mechanical principle that a local subsystem tends to maximize its entropy subject to global conservation laws (none, in this case). For the   projected ensemble, the entropy in question is that of the distribution of pure states over the Hilbert space. This condition can therefore be referred to as {\it deep thermalization}, as opposed to {\it regular thermalization}, which only constrains the behavior of ensemble-averaged observables, i.e., the distribution's mean.

That such universal distributions emerge was recently rigorously proven in a large family of models of quantum dynamics in one dimension~\cite{ho_exact_2022, ippoliti_dynamical_2022}.
Formally, the convergence to the uniform ensemble can be systematically probed by studying how well each statistical moment of the ensemble replicates that of the Haar distribution, a condition known in quantum information theory as forming an approximate {\it quantum state $k$-design}~\cite{ambainis_quantum_2007, gross_evenly_2007, low_pseudo-randomness_2010}, where integer $k\geq 1$ enumerates the statistical moments.
It is interesting, then, to ask about the timescales $t_k$ for which a projected ensemble forms a good $k$-design in quench dynamics (within some fixed accuracy, captured by an appropriate distance measure), and the physical processes which govern these time scales. 
Note that by construction, the {\it design times} $t_k$ are monotonically increasing ($t_{k+1}\geq t_k$), and that $t_1$ is the regular thermalization time.
Ref.~\cite{ho_exact_2022} found that $t_k = t_1$ for all $k$ in a Floquet Ising model tuned to a self-dual point, a result which was later extended to dual-unitary circuits\footnote{These are special circuits that are unitary not only in time, but also in space---a condition which makes these models analytically tractable in many cases~\cite{akila_particle-time_2016, bertini_exact_2018, bertini_exact_2019, gopalakrishnan_unitary_2019, claeys_maximum_2020}.} under more general conditions~\cite{claeys_emergent_2022, ippoliti_dynamical_2022}. Further, breaking of dual-unitarity in $(1+1)$-dimensional unitary circuits was shown to gap $t_\infty$ away from $t_1$ in general, via a mapping to measurement-induced dynamical purification~\cite{ippoliti_dynamical_2022}, which limits the transfer of information over long spatial distances in the system.
Aside from these results, our understanding of   how $t_k$ scales with $k$ and with other parameters of  dynamics in general is still rather limited. 

Here, we   consider a distinct and complementary mechanism that can gap $t_\infty$ away from $t_1$, having to do with {\it local} bottlenecks to the transfer of information in the system. 
To do so, we will focus on a model of chaotic quantum dynamics with minimal structure, meant to capture the basic feature of a spatially-local system: that the subsystem of interest $A$ interacts with the rest of the universe (an unstructured `bath') through the mediation of another subsystem (informally, its boundary). This minimal requirement is enough to give rise to an interesting separation between   regular and deep thermalization times, with the model remaining simple enough to allow for exact analytical results using random-matrix methods. 
 
We find that the projected ensemble on the subsystem of interest $A$, in the limit of an infinite random-matrix bath, almost always realizes the so-called {\it Scrooge}~\cite{jozsa_lower_1994, choi_emergent_2021} or {\it Gaussian Adjusted Projected} (GAP)~\cite{goldstein_distribution_2006, goldstein_universal_2016} ensemble---a maximally-entropic distribution of pure states consistent with a given density matrix, which is of independent interest in quantum information theory. We further compute statistical moments of this distribution and derive an exact form for the design times $t_k$ in our model. As a consequence of this result we show that (i) $t_k$ grows as $\log(k)$ for sufficiently small moments $k$, and (ii) this growth asymptotes at large $k$ to $t_\infty = 2t_1$.
This represents the second analytically-solved model of deep thermalization, after the dual-unitary circuit model of Ref.~\cite{ho_exact_2022}, and the first to exhibit a nontrivial separation between distinct design times $t_k$.
We conjecture that the latter phenomenology should also be present in systems in any dimension as long as they harbor local interactions, and whenever thermalization is {\it approximate}, i.e., when $\rho_A$ is close to, but not exactly equal to, a thermal equilibrium state (dual-unitary circuits are exceptional in this sense as thermalization there is exact~\cite{piroli_exact_2020}).
Thus, our findings provide a better understanding of the physical phenomenon of deep thermalization beyond the specific setting of the model studied in this work.

The paper is structured as follows.
In Sec.~\ref{sec:setup} we briefly review key concepts and introduce our model of dynamics.
An analytical solution for the design times is derived in Sec.~\ref{sec:derivation} and is tested against exact numerical simulations in Sec.~\ref{sec:numerics}. Finally, we summarize our results and discuss directions for future research in Sec.~\ref{sec:conclusion}.


\section{Setup \label{sec:setup}}

In this section, we first present a brief review of the object of interest, the projected ensemble, as well as the quantum information-theoretic formalism used to characterize it, namely quantum state designs (Sec.~\ref{sec:setup_review}).
For a more detailed review, see Refs.~\cite{cotler_emergent_2021, ippoliti_dynamical_2022}. 
We then define the specific model of dynamics considered in this work (Sec.~\ref{sec:setup_model}).

\subsection{The projected ensemble and quantum state designs \label{sec:setup_review}}

Consider a pure quantum state $\ket{\Psi}_{AB}$ on a bipartite system $AB$.
An ensemble $\mathcal{E}$ of pure states on $A$, called the projected ensemble\footnote{Such an ensemble is also an instance of a {\it geometric quantum state}~\cite{anza_beyond_2021, anza_quantum_2022}.}~\cite{cotler_emergent_2021}, can be naturally defined from it
via the outcomes of projective measurements   on subsystem $B$:
\begin{equation}
    \mathcal{E} = \{ (p(z), \ket{\psi_z}_A):\ z = 1, \dots d_B \} .
\end{equation}
Here $d_B$ is the Hilbert space dimension of $B$, $z$ labels an orthonormal basis\footnote{This is typically taken to be the computational basis in a system made of qubits. We do not make use of this structure in this paper.} of $B$, $p(z)$ is the probability of obtaining the outcome $z$ upon measuring $B$, and $\ket{\psi_z}_A$ is the post-measurement pure state of subsystem $A$, conditional on the outcome $z$. Explicitly, these are given by
\begin{equation}
    p(z) = \| \, {}_B\!\braket{z}{\Psi}_{AB} \|^2,
    \qquad 
    \ket{\psi_z}_A = {}_B\!\braket{z}{\Psi}_{AB} / \sqrt{p(z)}.
\end{equation}

If $\ket{\Psi}_{AB}$ is a typical state in the many-body Hilbert space $\mathcal{H}_A \otimes \mathcal{H}_B$, for sufficiently large $d_B$ the projected ensemble can be shown to acquire a universal form  with high probability, namely its states are distributed according to the Haar (i.e., unitarily-invariant) measure on $A$~\cite{cotler_emergent_2021}.
The same behavior was conjectured to emerge at late times in quantum chaotic many-body dynamics from simple initial states without conservation laws or at infinite temperature, hence representing a novel kind of nonequilibrium universality.
This was supported by extensive numerical results~\cite{cotler_emergent_2021}; later, rigorous exact results were obtained for dynamics in $(1+1)$-dimensional non-integrable dual-unitary circuits~\cite{ho_exact_2022, claeys_emergent_2022, ippoliti_dynamical_2022}. 
We note the projected ensemble is not only a theoretical construct, but is in fact also accessible in many present-day experimental platforms of programmable quantum simulators, owing to their ability to obtain site-resolved measurement data of the global system: by classically post-processing the data (for example, analyzing the distribution of bit-strings corresponding to measurement outcomes of $A$, correlated to a particular measurement outcome of $B$), one can probe the statistics of the projected ensemble~\cite{choi_emergent_2021}.

As mentioned in the introduction, the  approach of the distribution of states in the projected ensemble to the uniform distribution can be characterized in a systematic fashion: one can ask about the extent to which the projected ensemble $\mathcal{E}$ reproduces the $k$th moment of the an ensemble of states drawn from the Haar measure.
If the agreement is perfect, $\mathcal{E}$ is said to be a {\it quantum state $k$-design}, a concept from quantum information theory which has significant relevance in applications like randomized benchmarking and tomography~\cite{renes_symmetric_2004, ambainis_quantum_2007, knill_randomized_2008, roberts_chaos_2017, huang_predicting_2020, elben_randomized_2022}.
By allowing for a finite error $\epsilon >0$ in agreement (quantified by a distance measure, introduced below), this can be extended to the notion of an {\it approximate} quantum state $k$-design. 

Concretely, one can construct ``moment operators'' from $\mathcal{E}$, which are density matrices on a $k$-fold replicated Hilbert space $\mathcal{H}_A^{\otimes k}$, defined for all integer values of $k\geq 1$:
\begin{equation}
    \rho^{(k)} = \sum_z p(z) (\ket{\psi_z} \bra{\psi_z})^{\otimes k}
    \label{eq:moment_op}
\end{equation}
(we omit the subscript $A$ for brevity).
It is worth emphasizing that $\rho^{(k)} \neq \rho^{\otimes k}$ (i.e. the $k$-th moment operator for the ensemble $\mathcal{E}$ is \emph{not} equal to $k$ copies of the density matrix $\rho = \sum_z p(z) \ketbra{\psi_z}{\psi_z}$), and in fact $\rho^{(k)}$ is generally not a function of $\rho$ alone; in other words, the ensemble $\mathcal{E}$ genuinely contains more information than its associated density matrix $\rho$. 

For the particular ensemble of states distributed according to the Haar measure, the moment operators read
\begin{equation}
    \rho^{(k)}_H = \int {\rm d} \psi\ (\ket{\psi}\bra{\psi})^{\otimes k}
    = \binom{k+d_A-1}{k}^{-1} {\hat \Pi}_\text{symm}
    = \frac{(d_A-1)!}{(k+d_A-1)!} \sum_{\sigma \in S_k} \hat{\sigma} 
    \label{eq:haar_moment}
\end{equation}
where $\hat{\sigma}$ is an operator that shuffles the $k$ replicas of the Hilbert space according to a permutation $\sigma\in S_k$, and ${\hat \Pi}_\text{symm} = \frac{1}{k!} \sum_{\sigma\in S_k} \hat{\sigma}$ is the projector on the permutation-symmetric sector of $\mathcal{H}_A^{\otimes k}$, which has dimension $\binom{k+d_A-1}{k}$. 
The condition of $\mathcal{E}$ being a {\it quantum state $k$-design} is equivalent to the statement that $\rho^{(k)} = \rho_H^{(k)}$.

As the moment operator is a  density matrix, we can quantify the distance between two moment operators $\rho^{(k)}_1, \rho^{(k)}_2$ via standard distance measures for density matrices, like the trace norm $\frac{1}{2} \| \rho^{(k)}_1 - \rho^{(k)}_2 \|_1$. 
In this work, we will find it convenient to consider a distance measure based on the Frobenius norm
\begin{equation}
    \Delta^{(k)} 
    := \frac{\| \rho^{(k)} - \rho_H^{(k)}\|_2}{ \|\rho_H^{(k)} \|_2};
    \label{eq:distance_measure}
\end{equation}
 $\Delta^{(k)} = 0$ is then equivalent to $\mathcal{E}$ being a quantum state $k$-design, while $\Delta^{(k)} \leq \epsilon$ is then equivalent to $\mathcal{E}$ being an $\epsilon$-approximate quantum state $k$-design.
A closely related quantity is the {\it frame potential} of the ensemble, defined from the purity of the moment operators:
\begin{equation}
    F^{(k)} := \| \rho^{(k)} \|_2^2 
    = \text{Tr}[(\rho^{(k)})^2] 
    = \sum_{z_1,z_2} p(z_1) p(z_2) \left| \braket{\psi_{z_1} }{\psi_{z_2}} \right|^{2k}.
    \label{eq:frame_potential}
\end{equation}
For the Haar ensemble, the frame potential is $F_H^{(k)} = \binom{k+d_A-1}{k}^{-1}$. 
The distance measure $\Delta^{(k)}$, Eq.~\eqref{eq:distance_measure}, may be rewritten in terms of the frame potentials of the two ensembles (projected and Haar) as
\begin{equation}
    \Delta^{(k)} 
    = \left( \frac{F^{(k)}} {F_H^{(k)}} - 1\right)^{1/2}.
    \label{eq:distance_measure_vsF}
\end{equation}
This directly implies $F^{(k)} \geq F_H^{(k)}$, with equality if and only if the ensemble forms an exact state $k$-design: $\rho^{(k)} = \rho_H^{(k)}$.

Now, consider a quantum many-body state $|\Psi\rangle_{AB}$ which arises in quench dynamics, beginning from a disentangled state. 
We can define the time taken for its projected ensemble to form an $\epsilon$-approximate state $k$-design, i.e., the minimum $t$ such that $\Delta^{(k)} (t) \leq \epsilon$ for some arbitrary small threshold $\epsilon$. 
This defines the {\it design times} $t_k$ for each $k$. 
Due to the fact that $\Delta^{(k+1)} \geq \Delta^{(k)}$ as shown in Ref.~\cite{ippoliti_dynamical_2022}, these times are non-decreasing in $k$, $t_{k+1}\geq t_k$. 
For $k=1$ one recovers the `regular thermalization' time, since the first moment operator is simply the reduced density matrix of $A$: $\rho^{(1)} = \text{Tr}_B \ketbra{\Psi}{\Psi} = \rho_A$; thus $t_1$ is the time taken to achieve $\rho_A \simeq \rho_H^{(1)} = \mathbb{I}/d_A$, i.e., thermalization to infinite temperature\footnote{Here we implicitly assume either absence of any conservation laws, or an initial state whose expectation values of all conserved quantities match the infinite-temperature ones~\cite{srednicki_chaos_1994, rigol_thermalization_2008}.}.
The design times with $k>1$ may be viewed as probing   timescales associated with stronger forms of thermalization: equilibration not just of (averaged) expectation values, but also of statistical distributions of the conditional wavefunctions $|\psi_z\rangle$ themselves, which we may call {\it deep thermalization}.
In particular,  $t_\infty \equiv \lim_{k\to \infty} t_k$ (which exists due to monotonicity of $k$, though {\it a priori} may be infinite) is the time taken to completely reproduce the Haar measure.
Such deep thermalization is genuinely distinct from (and stronger than) conventional thermalization: it is fully possible to achieve the latter but not the former. As a paradigmatic example, a maximally-entangled ``EPR pair'' state $\ket{\Psi}_{AB} = \frac{1}{\sqrt{d_A}} \sum_{i=1}^{d_A}\ket{i}_A\otimes \ket{i}_B$ has an infinite-temperature reduced density matrix $\rho_A = \frac{1}{d_A} \sum_{i=1}^{d_A} \ketbra{i}{i}$, but its projected ensemble $\mathcal{E} = \{(p(z) = d_A^{-1}, \ket{z}_A):\ z=1,\dots d_A \}$ is very far from the Haar measure\footnote{It is easy to see that the ensemble's frame potential is $d_A^{-1}$, independent of $k$. Thus $\Delta^{(1)} = 0$ but e.g. $\Delta^{(2)} = \sqrt{(d_A-1)/2} \geq 1$.}.


\subsection{Model \label{sec:setup_model}} 

A key feature of any geometrically local, short-range interacting system (which many physical systems are) is that a contiguous subsystem $A$ exchanges information with the rest of the system through the mediation of other {\it local} subsystems. 
For example, a ball of radius $\xi$ in a $d$-dimensional lattice, $\{\mathbf{r} \in \mathbb{Z}^d:\ |\mathbf{r}|<\xi\}$, is surrounded by a shell $\{ \mathbf{r}\in\mathbb{Z}^d:\ \xi < |\mathbf{r}| < \xi'\}$ ($\xi'$ being another arbitrary radius $>\xi$) which contains a {\it finite} number of degrees of freedom\footnote{In fact, there are infinitely many such shells, with increasing radii $\xi', \xi'', \dots$, giving a hierarchy of potential bottlenecks that may prevent or slow down the transmission of information over long distances; Ref.~\cite{ippoliti_dynamical_2022} studies an example of this mechanism in $(1+1)$-dimensional quantum circuits, where the loss of information about measurement outcomes over long distances limits the effective size of the projected ensemble and slows down deep thermalization. Here instead we consider the effect of a single local bottleneck.}; with short-range interactions, any exchange of information with the outside world ($|\mathbf{r}|>\xi'$) must necessarily be mediated by this shell, which may act as a bottleneck and slow down the dynamics (compared e.g.~to a non-local, all-to-all interacting system).

\begin{figure}
    \centering
    \includegraphics[width=\textwidth]{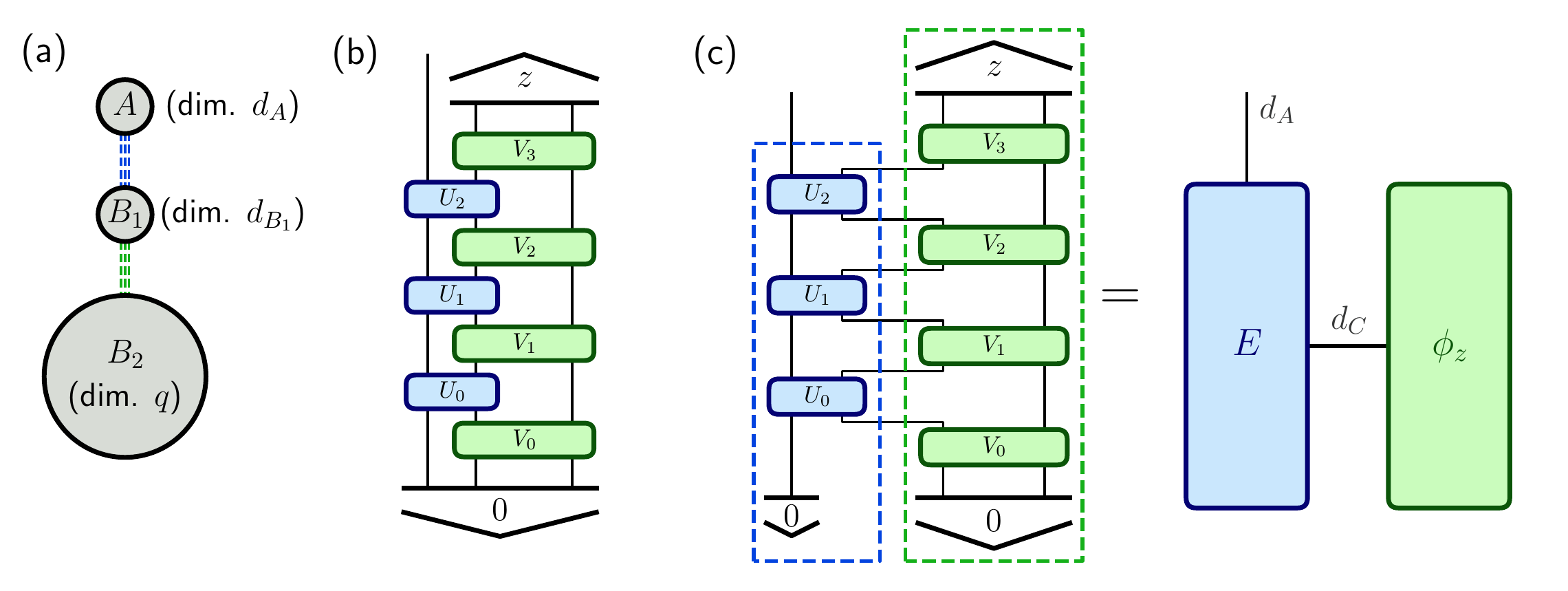}
    \caption{(a) A schematic of the model: a tripartite system $AB_1B_2$, with interactions between $A$ and $B_1$ (blue dashed lines) and between $B_1$ and $B_2$ (green dashed lines). The Hilbert space dimensions are also listed. 
    (b) Quantum circuit representing the dynamics of the model: an initial state $\ket{0}_{AB_1B_2}$ evolves under unitary gates $V_T \cdots U_0 V_0$ (here $T = 3$), then subsystem $B_1B_2$ is projectively measured in a fixed basis, yielding an outcome $z$ and a pure state $\ket{\psi_z}$ on $A$. 
    (c) The quantum circuit can be split into two blocks separated by a time-like cut that intersects the worldline of subsystem $B_1$ $2T$ times. The bond connecting the two blocks has dimension $d_C = d_{B_1}^{2T}$. One tensor can be viewed as a state $\ket{\phi_z}$ of subsystem $C$ (green box, right), while the other is a map $E$ from $C$ to $A$ (blue box, left). 
    We analyze the two parts in Sec.~\ref{sec:env} and \ref{sec:pe} respectively.
    }
    \label{fig:schematic}
\end{figure}

Here we consider a minimal example of a system featuring this mechanism, with a single ``local bottleneck'' mediating interactions between a finite subsystem of interest and the rest of the system, with the latter harboring nonlocal, all-to-all interactions. Our aim is to understand the effect of such locality constraints on the projected ensemble and on the emergence of higher state designs, i.e. deep thermalization. 
Concretely, we consider a bipartite system $AB$, where $B$ is the ``bath'' to be measured in order to generate the projected ensemble at $A$. In addition, we partition $B$ into two subsystems $B_1$ and $B_2$. $A$, $B_1$ and $B_2$ have Hilbert spaces of dimension $d_A$, $d_{B_1}$ and $q$, respectively. 
Subsystem $A$ interacts only with degrees of freedom in $B_1$, while $B_1$ and $B_2$ interact strongly and without spatial structure. 
This is shown schematically in Fig.~\ref{fig:schematic}(a). 

The system is initialized in a disentangled state $\ket{0}_A \ket{0}_{B_1} \ket{0}_{B_2}$ and subsequently evolves under a unitary circuit comprised of alternating gates on $AB_1$, $U_t\in U(d_Ad_{B_1})$, and on $B_1B_2$, $V_t\in U(d_{B_1} q)$. 
The gates $U_t$, $V_t$ are sampled from the Haar measure on the respective unitary groups (this means there is no additional locality structure inside the subsystems).
We define a time step to consist of a unitary $(U_t\otimes \mathbb{I}_{B_2})(\mathbb{I}_A \otimes V_t)$; the system is evolved for $T$ time steps ($t=0,\dots T-1$) and also by an additional gate $(\mathbb{I}_A \otimes V_T)$, following which $B$ is measured in an orthonormal basis $\{\ket{z}:\ z=1,\dots d_{B_1} q \}$, inducing a projected ensemble at $A$. This is illustrated in Fig.~\ref{fig:schematic}(b).

At this point it is worth remarking on different sources of randomness in the problem.
We wish to characterize the universal distributions on $A$ induced {\it solely} by randomness in the measurement outcomes at $B$ (that is, for a {\it fixed} realization of random gates $\{U_t, V_t\}$).
However, in our calculations we will utilize averages over the gates $\{V_t\}$. Note though that this is purely a technical tool to simplify calculations:  we will find that the behavior of the (gate-averaged) projected ensemble is {\it typical}, allowing us to make statements about the behavior of the projected ensemble for almost all circuit realizations.

Finally we remark on the structure of our toy model and how it relates to more realistic chaotic dynamics. 
Informally, subsystem $B_1$ plays the role of a local bottleneck that mediates correlations between $A$ and the rest of the universe, $B_2$, which is a large, unstructured random-matrix environment whose dimension $q$ will be taken to infinity. 
Clearly, this is a very favorable scenario for the generation of randomness on $A$: realistic local models in $d$-dimensional space will generically feature additional ``bottlenecks'' at all length scales, which we disregard here by taking a single bottleneck and an unstructured random environment.
For this reason, this serves as a useful minimal example of the effects of locality on deep thermalization: it is natural to conjecture that the exact results derivable in this model may turn into {\it bounds} for more realistic and structured models of dynamics.


\section{Analytical solution}
\label{sec:derivation}

In this Section we derive the main results of our work: an analytical computation of the distance measures $\Delta^{(k)}$ for the model of dynamics defined in Sec.~\ref{sec:setup_model}, leading to exact expressions for all design times and thus for deep thermalization, in the $q\to\infty$ limit. 
Our strategy is to divide the problem in two parts, as sketched in Fig.~\ref{fig:schematic}(c): first a treatment of the environment $B_2$ in the large-$q$ limit (Sec.~\ref{sec:env}) and then an analysis of subsystem $A$ through a replica method (Sec.~\ref{sec:pe}).
The results are then unpacked in Sec.~\ref{sec:designtimes}.

\subsection{Environment in the large-$q$ limit\label{sec:env}}

Here we show that, in the large-$q$ limit, subsystem $B$ (the ``environment'' with which subsystem $A$ interacts, and which is measured at the final time) can be replaced by a Haar-random state in the `bond space', i.e. the temporal cut\footnote{States defined on such temporal cuts in quantum circuits have been recently studied for various purposes, including simulation of quantum dynamics~\cite{banuls_matrix_2009,hastings_connecting_2015,lerose_influence_2021, sonner_influence_2021, giudice_temporal_2022}, realizations of monitored quantum circuits~\cite{ippoliti_postselection-free_2021, ippoliti_fractal_2022, lu_spacetime_2021}, measurement-based quantum computation~\cite{stephen_universal_2022} and quantum process tomography~\cite{pollock_non-markovian_2018}.} $C$ separating $A$ from $B$.

\begin{figure}
    \centering
    \includegraphics[width=1\textwidth]{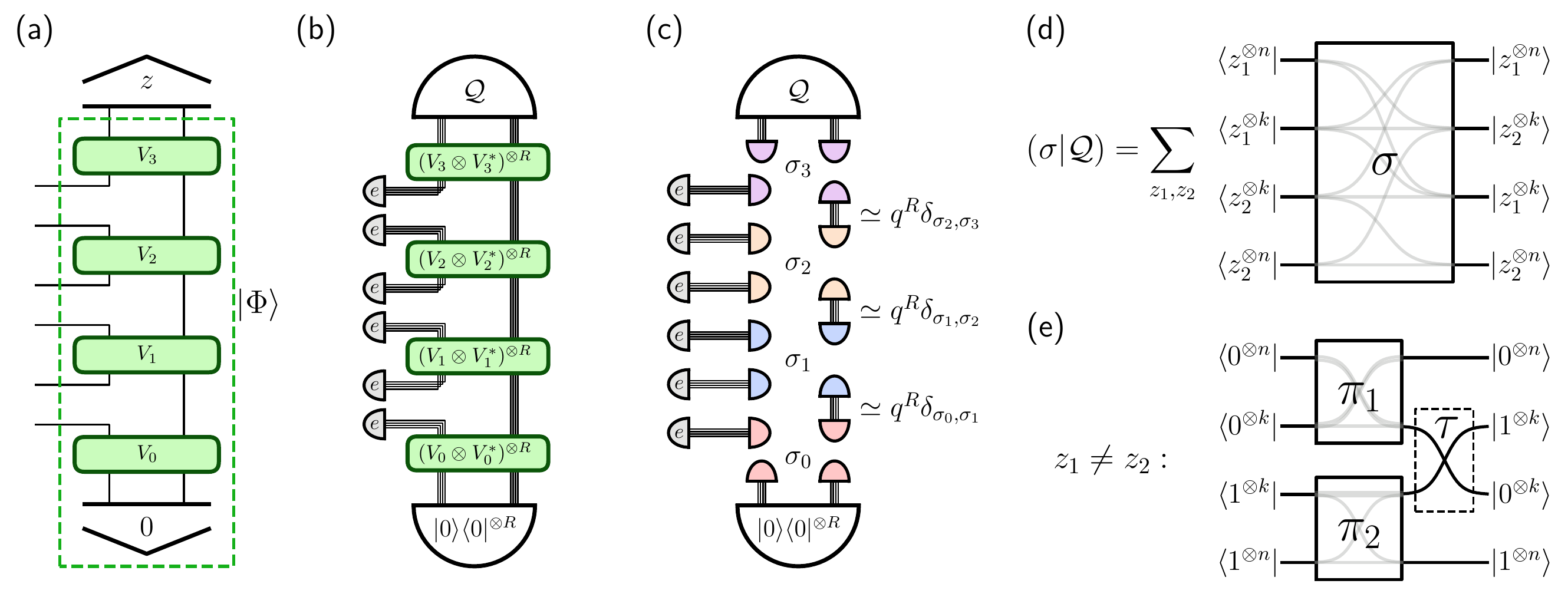}
    \caption{
    (a) Section of the quantum circuit in Fig.~\ref{fig:schematic}(c) representing the environment. The tensor enclosed in the green dashed box can be viewed as a state $\ket{\Phi}$ defined on systems $B$ (top legs) and $C$ (left legs). Upon projecting system $B$ on a basis state $\ket{z}$, one obtains  a state $\ket{\phi_z}$ on the bond space $C$, which has dimension $d_C = d_{B_1}^{2T}$ ($T=3$ in this example). 
    (b) Tensor network contraction representing the pseudo-frame potential $F^{(n,k)}$, Eq.~\eqref{eq:fnk}. $R = 2(n+k)$ is the total number of replicas. Semicircles represent operators acting on $R$ replicas of the Hilbert space, $e$ denotes the identity and $\mathcal{Q}$ is as in Eq.~\eqref{eq:Q}. 
    (c) The same tensor network contraction upon averaging $V_0,\dots V_T$ over the Haar measure on $U(d_{B_1} q)$, to leading order in $q\to\infty$. Averaging $(V_t\otimes V_t^\ast)^{\otimes R}$ yields a sum over permutations $\sigma_t \in S_R$, see Eq.~\eqref{eq:wg}; the contractions $(\sigma_t|\sigma_{t+1}) \propto \delta_{\sigma_t, \sigma_{t+1}}$ (to leading order in $q$) force all permutations $\sigma_t$ to coincide. 
    (d) Schematic of the inner product $(\sigma|\mathcal{Q})$.
    (e) For $z_1\neq z_2$ (as in Eq.~\eqref{eq:fnk_0011}), the inner product vanishes unless the permutation $\sigma$ takes the form $\tau \pi_1\pi_2$, with $\tau$ a transposition of the central $2k$ replicas as shown and $\pi_{1,2}\in S_{n+k}$ arbitrary.
    }
    \label{fig:env_diagrams}
\end{figure}

The circuit setup is illustrated in Fig.~\ref{fig:env_diagrams}(a), where the environment ($B$, comprised of $B_1$ and $B_2$, with dimensions $d_{B_1}$ and $q$ respectively, see Sec.~\ref{sec:setup}) is initialized in $\ket{0}$ and measured at the end to find computational basis state $\ket{z}$ ($z \in \{1, \dots d_{B_1} q\}$). 
This gives a state $\ket{\phi_z}$ at the timelike subsystem $C$, which is made of $2T$ qudits of dimension $d_{B_1}$ each ($T$ is the circuit depth as defined earlier), giving a total dimension $d_C = d_{B_1}^{2T}$. 
We may view this state as belonging to  a projected ensemble on $C$ from a bipartite state $\ket{\Phi}_{BC}$, where $B$ is being measured. 
Concretely, let us introduce the unnormalized states $| \tilde{\phi}_z \rangle_C \equiv {}_B\! \braket{z}{\Phi}_{BC}$, with $\ket{\Phi}$ the tensor in Fig.~\ref{fig:env_diagrams}(a), which obeys\footnote{This is seen by contracting the corresponding tensor network from the final time backwards, using unitarity of the $V_t$ gates; this produces a factor of $d_{B_1}$ per time step, so overall $d_{B_1}^T = d_C^{1/2}$.} $\braket{\Phi}{\Phi} = d_C^{1/2}$.
Then, the ensemble at $C$ is given by 
\begin{equation} 
\mathcal{E}_C = \left\{ \left( 
p(z) = \frac{ \langle \tilde{\phi}_z | \tilde{\phi}_z \rangle}{d_C^{1/2}} , 
\ket{\phi_z} = \frac{ |\tilde{\phi}_z\rangle}{ \| |\tilde{\phi}_z\rangle \|}
\right):\ z = 1, \dots d_{B_1}q \right\}. \label{eq:pe_C}
\end{equation}
Note the normalization of $p(z)$ follows from $\sum_z p(z) = d_C^{-1/2} \braket{\Phi}{\Phi} = 1$.

In what follows, we are going to show that for $q\to\infty$, $\mathcal{E}_C$ approaches an ensemble of states distributed according to the Haar measure
on $\mathcal{H}_C$.
To do so, we aim to compute the frame potential:
\begin{equation}
    F^{(k)} = \sum_{z_1, z_2} p(z_1) p(z_2) \left|\braket{\phi_{z_1}}{\phi_{z_2}} \right|^{2k}
    = \frac{1}{d_C} \sum_{z_1, z_2} 
    \braket{\tilde{\phi}_{z_1}}{\tilde{\phi}_{z_1}}^{1-k} 
    \left| \braket{\tilde{\phi}_{z_1}}{\tilde{\phi}_{z_2}} \right|^{2k}
    \braket{\tilde{\phi}_{z_2}}{\tilde{\phi}_{z_2}}^{1-k}.
\end{equation}  

To make progress, we use a replica technique~\cite{claeys_emergent_2022} to define a quantity $F^{(n,k)}$ which reduces to the frame potential $F^{(k)}$ upon taking the ``replica limit'' $n \to 1-k$:
\begin{align}
    F^{(n,k)} 
    & = \frac{1}{d_C} \sum_{z_1,z_2} \braket{\tilde{\phi}_{z_1} }{ \tilde{\phi}_{z_1} }^n 
    \left| \braket{\tilde{\phi}_{z_1} }{ \tilde{\phi}_{z_2} }  \right|^{2k}
    \braket{\tilde{\phi}_{z_2} }{ \tilde{\phi}_{z_2} }^n
    = \frac{1}{d_C} \Tr \left( \ketbra{\Phi}{\Phi}^{\otimes R}_{BC} \mathcal{Q}_B \right)
    \label{eq:fnk} \\
    \mathcal{Q}_B
    & = \sum_{z_1, z_2} \ket{z_1^{\otimes n} z_1^{\otimes k} z_2^{\otimes k} z_2^{\otimes n}}_B
    \bra{z_1^{\otimes n} z_2^{\otimes k} z_1^{\otimes k} z_2^{\otimes n}}_B
    \label{eq:Q}
\end{align}
where $R = 2(n+k)$ denotes the total number of replicas.
The $F^{(n,k)}$ quantity, represented as a tensor network contraction in Fig.~\ref{fig:env_diagrams}(b), involves replicated Haar-random unitaries $(V_t \otimes V_t^\ast)^{\otimes R}$. 
These tensor products can be averaged exactly~\cite{kostenberger_weingarten_2021}, and to leading order in $q\to\infty$ one gets
\begin{equation}
    \mathbb{E}_{V_t \sim \text{Haar}}[(V_t \otimes V_t^\ast)^{\otimes R} ]
    = \sum_{\sigma, \pi \in S_R } \text{Wg}_{d_{B_1} q} (\sigma^{-1}\pi, R) |\hat \sigma) (\hat \pi|
    \simeq \text{Wg}_{d_{B_1} q} (e, R) \sum_{\sigma \in S_R } |\hat \sigma) (\hat \sigma|
    \label{eq:wg}
\end{equation}
where $\text{Wg}_d(\sigma,R)$ is the Weingarten function~\cite{kostenberger_weingarten_2021} for $R$ replicas of a $d$-dimensional Hilbert space evaluated on a permutation $\sigma\in S_R$, 
$e\in S_R$ is the identity permutation, and $\hat{\sigma}$ represents an operator that shuffles the $R$ replicas according to a permutation $\sigma$: $\hat{\sigma}\ket{i_1,\dots i_R} = \ket{i_{\sigma^{-1}(1)}, \dots i_{\sigma^{-1}(R)}}$. 
Additionally, here and in the following we use a standard operator-state mapping notation where an operator $O = \sum_{a,b} O_{ab}\ketbra{a}{b}$ defines a state $|O) = \sum_{ab} O_{ab} \ket{a}\otimes\ket{b}$ on a doubled Hilbert space, with inner product $(O|O') = \text{Tr}(O^\dagger O')$. Thus e.g. $|\hat{\sigma})(\hat{\sigma}|$ in Eq.~\eqref{eq:wg} is a super-operator whose action is defined by $O \mapsto \hat{\sigma} {\rm Tr}(\hat{\sigma}^\dagger O)$.

The computation of $F^{(n,k)}$, Eq.~\eqref{eq:fnk}, features $T+1$ uncorrelated gates $V_t$, all of which are independently averaged according to Eq.~\eqref{eq:wg}. We have that $\text{Wg}_{d_{B_1} q}(e,R) = (d_{B_1} q)^{-R}$ to leading order in large $q$, contributing an overall prefactor of $(d_{B_1} q)^{-R(T+1)}$.
In addition, the averaging produces $T$ inner products between permutations (acting on $B_2$ only), as sketched in Fig.~\ref{fig:env_diagrams}(c).
We have $(\hat \sigma| \hat \pi) = q^{|\sigma^{-1}\pi|}$ where $|\cdots|$ denotes the number of cycles in a permutation. 
To leading order in large $q$, this forces all permutations to be the same: if $\sigma = \pi$ we have $q^{|\sigma^{-1}\pi|} = q^{|e|} = q^R$; otherwise $\sigma^{-1}\pi$ has strictly fewer than $R$ cycles giving a subleading contribution. 
Combining all these inner products yields a factor of $q^{RT}$. 

In all, we obtain 
\begin{align}
    \mathbb{E}[ F^{(n,k)} ] 
    & \simeq d_C^{-1} d_{B_1}^{-R(T+1)} q^{-R} \sum_{\sigma \in S_R} 
    \sum_{z_1, z_2 = 1}^{d_{B_1} q}
    \bra{0^{\otimes R}} \hat{\sigma} \ket{0^{\otimes R}}_B 
    \text{Tr}_C(\hat\sigma)
    \text{Tr}_B(\hat\sigma \mathcal{Q})
\end{align}
where the three factors inside the summation arise from contraction of $\hat{\sigma}$ with the circuit's boundary conditions at the initial time ($\ket{0}$), final time (operator $\mathcal{Q}$ from Eq.~\eqref{eq:Q}, representing projection on all possible $z_1$, $z_2$ outcomes on various replicas), and spatial edge (tracing out $C$) respectively, as illustrated in Fig.~\ref{fig:env_diagrams}(c).
For the first factor, we note that (as $\ket{0}^{\otimes R}$ is permutation-invariant) $\bra{0^{\otimes R}} \hat{\sigma} \ket{0^{\otimes R}} = \braket{0}{0}^R = 1$. 
The second factor is sketched diagrammatically in Fig.~\ref{fig:env_diagrams}(d). The summation over $z_{1,2}$ (see Eq.~\eqref{eq:Q}) breaks up into two pieces: 
$d_{B_1} q$ terms with $z_1 = z_2$, and $(d_{B_1}q)^2 - d_{B_1}q$ terms with $z_1\neq z_2$. 
There is no further dependence on $q$, so in the $q\to\infty$ limit we can focus on the latter terms. 
We get
\begin{equation}
    \mathbb{E}[ F^{(n,k)} ] 
    \simeq d_C^{-1-R/2} (d_{B_1} q)^{2-R} \sum_{\sigma \in S_R}
    \bra{0^{\otimes n} 0^{\otimes k} 1^{\otimes k} 1^{\otimes n}} \hat \sigma \ket{0^{\otimes n} 1^{\otimes k} 0^{\otimes k} 1^{\otimes n}}_B 
    \text{Tr}_C \hat\sigma 
    \label{eq:fnk_0011}
\end{equation}
where we have inserted 0 and 1 as arbitrary (orthogonal) computational basis states on $B$. 
The term $\bra{0^{\otimes n} 1^{\otimes k} 1^{\otimes k} 0^{\otimes n}} \hat \sigma \ket{0^{\otimes n} 1^{\otimes k} 0^{\otimes k} 1^{\otimes n}}$ serves to restrict the summation to permutations in the form $\sigma = \tau \pi_1 \pi_2$, where $\tau$ is a transposition of the central $2k$ replicas, while $\pi_{1,2}$ act on the first and last $n+k$ replicas, respectively, as illustrated in Fig.~\ref{fig:env_diagrams}(e). 
One can see that any permutation not in this form would produce an inner product $\braket{0}{1} = 0$.

In conclusion we arrive at 
\begin{align}
    \mathbb{E}[ F^{(n,k)} ] 
    & \simeq d_C^{-1-R/2} (d_{B_1}q)^{2-R} \sum_{\pi_{1,2} \in S_{n+k}} \text{Tr}[\hat{\tau}\cdot(\hat{\pi}_1\otimes \hat{\pi}_2)] \nonumber \\
    & = d_C^{-1-R/2} (d_{B_1}q)^{2-R} \left[\frac{(R/2+d_C-1)!}{(d_C-1)!}\right]^2 
    \text{Tr}[\hat\tau  \cdot (\rho_{H,C}^{(n+k)} \otimes \rho_{H,C}^{(n+k)})] \nonumber \\
    & = d_C^{-1-R/2} (d_{B_1}q)^{2-R}  \left[\frac{(R/2+d_C-1)!}{(d_C-1)!}\right]^2 F_H^{(k)}
\end{align}
where in the last line we have traced out the $2n$ replicas (on which the transposition $\hat{\tau}$ does not act) obtaining $(\rho_{H,C}^{(k)})^{\otimes 2}$, and then noted that $\text{Tr}(\hat{\tau} A^{\otimes 2}) = \text{Tr}(A^2)$.
Finally, having obtained an analytical function of $R$, we continue it to $R = 2(n+k) = 2$ (i.e.~$n=1-k$), which gives the desired result:
\begin{equation}
    \lim_{q\to\infty} \mathbb{E}[ F^{(k)} ] = F_H^{(k)}.
\end{equation}
Given that $F^{(k)} \geq F_H^{(k)}$ in each realization (following Eq.~\eqref{eq:distance_measure_vsF}), we can conclude that $F^{(k)} = F_H^{(k)}$ {\it almost always} (over the Haar measure for gates $V_t$). 
This means that statistically speaking, we can replace the entire subsystem $B$ with a Haar-random state on the temporal cut $C$ in order to analyze the formation of state designs on $A$.

Some comments are in order here.
First, we note that the same result holds in dual-unitary circuits with suitable initial states and final measurement bases in 1+1 dimensions~\cite{ho_exact_2022, claeys_emergent_2022, ippoliti_dynamical_2022}: there, too, one obtains the Haar distribution in the bond space $C$ in the limit of an infinite bath, $N_B \to \infty$ where $N_B$ is the number of degrees of freedom in $B$ (the corresponding thermodynamic limit in our case is $q\to\infty$). 
In that case, the Haar distribution emerges essentially from a deep random unitary circuit running in the space direction; here instead it emerges from measuring part of a given highly-entangled state $\ket{\Phi}_{BC}$, similarly to the case of a global Haar-random state $\ket{\Psi}_{AB}$ studied in Ref.~\cite{cotler_emergent_2021}.
Secondly, we note that realistic models of dynamics may fail to produce Haar-randomness in the bond space ($C$ in our case); this is indeed the subject of Ref.~\cite{ippoliti_dynamical_2022}, where it was shown that $(1+1)$-dimensional unitary circuits may fail to achieve this due to dynamical purification in the space direction, i.e., the loss of information about measurement outcomes over very long distances. Here, by taking an all-to-all interacting random-matrix environment, we have avoided any such issues. Therefore this is a favorable scenario for the formation of state designs at $A$, and we expect that results obtained in this model may provide bounds for more realistic short-range-interacting models.

\subsection{Projected ensemble on $A$ \label{sec:pe}}

With the  above result we can greatly simplify the setup, by discarding the thermodynamically-large environment $B_2$ and focusing only on subsystem $A$ and the temporal subsystem (or bond space) $C$, as sketched in Fig.~\ref{fig:schematic}(c).
After observing outcome $z$ on $B$, we have a state $\ket{{\phi}_z}$ at $C$ which gets mapped to $A$ by a linear map\footnote{The map $E$ is obtained by combining the gates $\{U_t\}$ as shown diagrammatically in Fig.~\ref{fig:schematic}(c). Formally, one has $E_{a,\mathbf{b}} = \sum_{\alpha_0, \dots \alpha_T} \delta_{\alpha_0, 0} \delta_{\alpha_T,a} \prod_{t=0}^{T-1} \bra{\alpha_{t+1},b_{2t+1}} U_t \ket{\alpha_t, b_{2t}}$, where $a$ and each $\alpha_t$ run over states of $A$, each $b_t$ runs over states of $B_1$, and the multi-index $\mathbf{b} = (b_0,\dots b_{2T-1})$ runs over states of $C$.} $E$.
As we just saw, in the large-$q$ (thermodynamic) limit the input state $\ket{\phi_z}$ is effectively Haar-random on $C$, thus the question becomes: how well is this randomness propagated from $C$ to $A$? 

In the case of dual-unitary circuits in $1+1$ dimensions, the randomness is propagated optimally: as long as $d_C\geq d_A$, the Haar measure at $C$ induces the Haar measure at $A$.
This follows from the {\it perfect thermalization} of dual-unitary circuits~\cite{piroli_exact_2020}: because $\rho_A = \mathbb{I}_A / d_A$, $E$ must act as an isometry on a subspace of $\mathcal{H}_C$ (of the appropriate dimension $d_A$), which is enough to produce the Haar measure on $\mathcal{H}_A$~\cite{ho_exact_2022}. 
However, for generic interactions (i.e. non-dual-unitary, or in higher dimension, etc), this is not the case: $\rho_A$ equals $\mathbb{I}/d_A$ only approximately, and it is not {\it a priori} clear what the induced distribution of states at $A$ looks like. This is the question we address next.

In terms of the states $|\tilde{\phi}_z \rangle_C = {}_B\! \braket{z}{\Phi}_{BC}$ studied in Sec.~\ref{sec:env}, the projected ensemble at $A$ takes the form
\begin{equation}
    \mathcal{E} = \left\{ \left( p(z) =  
    \langle \tilde{\phi}_z |E^\dagger E | \tilde{\phi}_z \rangle, \ 
    \ket{\psi_z} = \frac{E | \tilde{\phi}_z \rangle }{\| E | \tilde{\phi}_z \rangle \|} 
    \right):\ z = 1,\dots d_{B_1}q \right\}
    \label{eq:finite_q_distribution}
\end{equation}
It follows that the expectation of a function $f(\ket{\psi})$ on the ensemble $\mathcal{E}$ is
\begin{equation}
    \mathbb{E}f = \sum_z p(z) f(\ket{\psi_z}) = \sum_z \braket{\tilde{\phi}_z}{\tilde{\phi}_z} 
    \bra{\phi_z} E^\dagger E \ket{\phi_z} f\left( \frac{E\ket{\phi_z}}{ \| E \ket{\phi_z} \| } \right) \label{eq:example_averaging}
\end{equation}
where we have introduced the normalized vectors $\ket{\phi_z} = |\tilde{\phi}_z\rangle / \| |\tilde{\phi}_z\rangle  \|$. 
The result of Sec.~\ref{sec:env} is that the ensemble in Eq.~\eqref{eq:pe_C} is Haar-distributed in the thermodynamic limit $q\to\infty$, i.e.,
\begin{equation}
    \lim_{q\to\infty} \sum_z d_C^{-1/2} \braket{\tilde{\phi}_z}{\tilde{\phi}_z} g(\ket{\phi_z}) = \int {\rm d}\phi\ g(\ket{\phi})
\end{equation}
with ${\rm d}\phi$ the Haar measure and $g$ any sufficiently regular function.
With this substitution, Eq.~\eqref{eq:example_averaging} becomes
\begin{equation}
    \mathbb{E}f = d_C^{1/2} \int {\rm d}\phi\ \bra{\phi}E^\dagger E \ket{\phi} f \left( \frac{E\ket{\phi}}{\| E\ket{\phi} \| } \right)
\end{equation}
which defines the limiting form of the projected ensemble at $A$:
\begin{equation}
    \lim_{q\to\infty} \mathcal{E} = \left\{ \left({\rm d}\psi = {\rm d}\phi\ d_C^{1/2} \bra{\phi}E^\dagger E\ket{\phi},\ \ket{\psi} = \frac{E\ket{\phi}}{\| E\ket{\phi}\|} \right) \right\},
    \label{eqn:limiting_distribution}
\end{equation}
where again ${\rm d}\phi$ is the Haar measure. 

The resulting ensemble, Eq.~\eqref{eqn:limiting_distribution}, is known as the {\it Scrooge ensemble}~\cite{jozsa_lower_1994, choi_emergent_2021} or {\it Gaussian Adjusted Projected} (GAP) {\it ensemble}~\cite{goldstein_distribution_2006, goldstein_universal_2016}, which is the maximally entropic\footnote{More precisely, it minimizes a measure of accessible information~\cite{jozsa_lower_1994}, thus it is ``maximally stingy'' with its information, hence the name.} ensemble of pure states compatible with a given reduced density matrix---in this case,
\begin{equation}
    \rho_A = \int {\rm d}\psi\ \ketbra{\psi}{\psi}
    = \int {\rm d}\phi\ d_C^{1/2} E\ketbra{\phi}{\phi} E^\dagger 
    = d_C^{-1/2} EE^\dagger. \label{eq:rdm}
\end{equation}
As a technical remark, we note that Eq.~\eqref{eqn:limiting_distribution} is seemingly different from the conventional construction of the Scrooge/GAP ensemble (as presented in \cite{jozsa_lower_1994, choi_emergent_2021}), as the former involves objects acting on two Hilbert spaces of different dimensions ($A$ and $C$) while the latter only one. However, it can be straightforwardly shown that the two definitions are equivalent, see Appendix~\ref{app:scrooge}.

\subsection{Moments of the Scrooge/GAP ensemble near infinite temperature \label{sec:scrooge}}

To recapitulate, the analysis of Sec.~\ref{sec:env} and \ref{sec:pe} has shown that, in the large-$q$ (thermodynamic) limit, the projected ensemble constructed from our model of dynamics realizes (almost always) the Scrooge/GAP ensemble, Eq.~\eqref{eqn:limiting_distribution}, induced by the appropriate density matrix, Eq.~\eqref{eq:rdm}. 
With this result in hand, we are now in a position to study the formation of higher state designs at $A$. 
However, we remark that the derivation which follows is more general: it may be applied to the Scrooge/GAP ensemble built from any density matrix $\rho$ on any system, as long as $\rho$ is very close to thermalization at infinite temperature. Such an ensemble may potentially be obtained through other means (i.e., not by the projected ensemble on our model) and is an object of independent interest.

In order to characterize the emergence of quantum state designs, we aim to compute the $k$th moment operators for the ensemble in Eq.~\eqref{eqn:limiting_distribution}:
\begin{align}
    \rho_A^{(k)} 
    & = \int {\rm d}\psi (\ketbra{\psi}{\psi})^{\otimes k}
    = d_C^{1/2} \int {\rm d}\phi\ \| E\ket{\phi} \|^2 \left( \frac{E\ketbra{\phi}{\phi}E^\dagger}{\| E\ket{\phi}\|^2 } \right)^{\otimes k} \nonumber \\
    & = d_C^{1/2} \int {\rm d}\phi\ \bra{\phi} E^\dagger E \ket{\phi}^{1-k} (E\ketbra{\phi}{\phi} E^\dagger)^{\otimes k}
    \label{eq:rhoA_k}.
\end{align}
At this point, for $k>1$ the presence of a negative power $1-k$ prevents us from directly performing the integral over the Haar measure ${\rm d}\phi$. 
This issue can be overcome by using a replica trick~\cite{claeys_emergent_2022} analogous to the one in Eq.~\eqref{eq:fnk}: we define a ``pseudo-moment operator''
\begin{equation}
    \rho^{(n,k)}_A = d_C^{1/2}  \int {\rm d}\phi\ \bra{\phi}E^\dagger E \ket{\phi}^n  (E\ket{\phi}\bra{\phi}E^\dagger)^{\otimes k}
\end{equation}
which reduces to $\rho^{(k)}_A$ for $n = 1-k$. We take $n>0$ in order to perform the calculation; at the end we will take the ``replica limit'' $n \to 1-k$ and obtain the desired result.
By using the fact that $\braket{\phi}{\phi} = \text{Tr} \ketbra{\phi}{\phi}$, we may re-write this as
\begin{align}
\rho^{(n,k)}_A
& = d_C^{1/2} \Tr_{A_1,\dots A_n} \int {\rm d} \phi\ 
(E\ket{\phi}\bra{\phi}E^\dagger)^{\otimes (n+k)} 
\end{align}
where we have introduced $n$ new replicas of $A$ in addition to the $k$ replicas that go into the definition of the $k$-th moment operator, and the trace is taken over these new replicas $A_1, \dots A_n$.

We can now perform the integral over the Haar measure ${\rm d} \phi$ exactly:
\begin{align} 
\rho^{(n,k)}_A 
& = d_C^{1/2}  \Tr_{A_1, \dots A_n}\left[ E^{\otimes(n+k)} \rho_{H,C}^{(n+k)} (E^\dagger)^{\otimes (n+k)}\right] \nonumber \\
& = d_C^{1/2}  \frac{(d_C-1)!}{(d_A-1)!} \frac{(n+k+d_A-1)!}{(n+k+d_C-1)!} \Tr_{A_1, \dots A_n} \left[\rho_{H,A}^{(n+k)} (EE^\dagger)^{\otimes (n+k)}\right]
\label{eq:pseudomoment_vs_ee}
\end{align}
where $\rho_{H,A}^{(n+k)}$ and $\rho_{H,C}^{(n+k)}$ are the moment operators of the Haar ensemble on $A$ and $C$, respectively.
In the second line we have commuted $E^{\otimes(n+k)}$ past $\rho_{H,C}^{(n+k)}$ by using the decomposition of the Haar moment operators into permutations and the fact that $E^{\otimes (n+k)}$ is permutation-symmetric: for any integer $m>0$, we have
\begin{align}
    E^{\otimes m} \rho_{H,C}^{(m)}
    & = \frac{(d_C-1)!}{(m+d_C-1)!} \sum_{\sigma \in S_m} E^{\otimes m} \hat{\sigma}_C
    = \frac{(d_C-1)!}{(m+d_C-1)!} \sum_{\sigma \in S_m} \hat{\sigma}_A (\hat{\sigma}_A^{-1} E^{\otimes m} \hat{\sigma}_C) \nonumber \\
    & = \frac{(d_C-1)!}{(m+d_C-1)!} \frac{(m+d_A-1)!}{(d_A-1)!} \rho_{H,A}^{(m)} E^{\otimes m}.
\end{align}
Next, we use Eq.~\eqref{eq:rdm} to eliminate $EE^\dagger$ from Eq.~\eqref{eq:pseudomoment_vs_ee}, obtaining
\begin{align}
\rho^{(n,k)}_A
& = g(n,k,d_A,d_C) \Tr_{A_1, \dots A_n}[\rho_{H,A}^{(n+k)} (d_A \rho_A)^{\otimes (n+k)}] \;,
\label{eq:pseudomoment_vs_rhoA}
\end{align}
where we have collected the awkward prefactors into a function $g$ which can be seen to equal 1 for all $d_A$, $d_C$ when $n+k=1$, and can thus be safely dropped for our purpose.

The expression Eq.~\eqref{eq:pseudomoment_vs_rhoA} is noteworthy as it only involves the reduced density matrix $\rho_A$ and the moment operator of the Haar ensemble. It is clear that if $\rho_A = \mathbb{I}/d_A$ (perfect thermalization), then $\rho_A^{(n,k)} = \text{Tr}_{A_1, \dots A_n}[ \rho_{H,A}^{(n+k)}] = \rho_{H,A}^{(k)}$, independent of $n$, and we have an exact quantum state $k$-design for all $k$, i.e., the exact Haar measure at $A$. 
Conversely, any deviation $\rho_A\neq \mathbb{I}/d_A$ (imperfect thermalization) directly maps onto a deviation between the $k$th moment operators. 
This deviation is what we aim to quantify next.

Let us assume that $\rho_A^{(1)} = (\mathbb{I} + \mu)/d_A$, where $\mu$ is a Hermitian operator that represents the slowest observable in the process of relaxation towards equilibrium at infinite temperature. 
By definition, we have 
\begin{equation}
    \Delta^{(1)} = \frac{\| \rho_A -\mathbb{I}/d_A\|_2 }{\|\mathbb{I}/d_A\|_2 } = \frac{\| \mu \|_2}{ \| \mathbb{I} \|_2}. 
    \label{eq:delta_1}
\end{equation}
At late times, when $\| \mu \|_2 $ is small, we can expand Eq.~\eqref{eq:pseudomoment_vs_rhoA} in powers of $\mu$:
\begin{align}
\rho^{(n,k)}_A
& = \rho_{A,H}^{(k)} + \Tr_{A_1, \dots A_n} \left[\rho_{H,A}^{(n+k)} \sum_{j=1}^{n+k} \mu_j \right] + O(\mu^2),
\end{align}
where $\mu_j$ denotes a $\mu$ operator acting on replica $A_j$ only (with $\mathbb{I}$ on all other replicas). 
Focusing on the linear term, we can split the sum into two parts:
\begin{itemize}
    \item Terms with $j>n$: tracing out $A_1, \dots A_n$ directly gives $\rho_{H,A}^{(k)} \sum_{j=1}^k \mu_j$;
    \item Terms with $1\leq j \leq n$: due to permutation symmetry these $n$ terms are all identical, giving $n$ contributions of $\text{Tr}_{A_{k+1}}[\rho_{H,A}^{(k+1)} \mu_{k+1}]$. 
\end{itemize}
The latter can be computed as follows. Decomposing $\rho_{H,A}^{(k+1)}$ into permutations $\sigma \in S_{k+1}$, we find that permutations with $\sigma(k+1) = k+1$ give a contribution $\propto {\rm Tr}(\mu) = 0$ ($\mu$ is traceless by definition). On the other hand, permutations with $\sigma(k+1) = j\leq k$ give a contribution $\propto \mu_j \hat{\pi}$, where $\pi\in S_k$ is a permutation induced on the remaining $k$ replicas\footnote{There is a one-to-one correspondence between $\pi\in S_k$ and $\sigma\in S_{k+1}$ with the constraint $\sigma(k+1)=j$.}. 
We have
\begin{align}
    {\rm Tr}_{A_{k+1}} \rho_{H,A}^{(k+1)} \mu_{k+1} 
    & = \frac{(d_A-1)!}{(k+d_A)!} \sum_{j=1}^k 
    \sum_{\substack{ \sigma\in S_{k+1}:\\ \sigma(k+1)=j}} {\rm Tr}_{A_{k+1}} \hat{\sigma} \mu_{k+1} \nonumber \\
    & = \frac{(d_A-1)!}{(k+d_A)!} \sum_{j=1}^k \mu_j \sum_{\pi\in S_k} \hat{\pi} 
    = \frac{1}{k+d_A} \sum_{j=1}^k \mu_j \rho_{H,A}^{(k)}.
\end{align}
Adding both contributions, we conclude
\begin{align}
\rho_{A}^{(n,k)} - \rho_{H,A}^{(k)} & = \left( 1 + \frac{n}{d_A+k} \right) \rho_{H,A}^{(k)} \sum_{j=1}^k \mu_j + O(\mu^2).
\end{align}
At this point, all the $n$ auxiliary replicas have been traced out, and the parameter $n$ appears only as a prefactor.
It is possible to take the replica limit\footnote{Strictly speaking, this assumes that the operations of replica limit and Taylor expansion to $O(\mu)$ commute. As we will see in Sec.~\ref{sec:numerics}, this assumption is justified by excellent agreement with numerical simulations.} $n \to 1-k$, which gives
\begin{equation}
\rho_A^{(k)} - \rho_{H,A}^{(k)} = \frac{d_A+1}{d_A+k} \rho_{H,A}^{(k)} \sum_{j=1}^k \mu_j + O(\mu^2).
\end{equation}
This is one of the central results of our work, directly connecting imperfect thermalization (quantified by $\mu$) to the discrepancy between $k$-th moments of the projected and Haar ensembles.

With this result in hand, we may now evaluate the distance measure Eq.~\eqref{eq:distance_measure}:
\begin{align}
(\Delta^{(k)})^2 
& = \frac{\text{Tr}[( \rho_A^{(k)} - \rho_{A.H}^{(k)})^2 ]}{\text{Tr}[(\rho_{H,A}^{(k)})^2 ]}
= \left( \frac{d_A+1}{d_A+k} \right)^2 \Tr[ \rho_{H,A}^{(k)} \sum_{i,j=1}^k \mu_i \mu_j]
\end{align}
where we used the facts that $\sum_i \mu_i$ commutes with $\rho_{H,A}^{(k)}$ (being permutation-symmetric) and that $(\rho_{H,A}^{(k)})^2 / \text{Tr}[(\rho_{H,A}^{(k)})^2] = \rho_{H,A}^{(k)}$ in order to simplify the result.
The summation over replicas $i$, $j$ breaks up into two parts:
\begin{itemize}
    \item Terms $i=j$ give $k$ identical contributions of $\text{Tr}(\rho_{H,A}^{(1)} \mu_1^2) = \| \mu \|_2^2/ d_A$. 
    \item Terms $i\neq j$ give $k(k-1)$ identical contributions of $\text{Tr}( \rho_{H,A}^{(2)} \mu_1 \mu_2)$; using $\rho_{H,A}^{(2)} = \frac{\hat{e} + \hat{\tau}}{d_A(d_A+1)}$ (where $e, \tau \in S_2$ are the identity permutation and the transposition, respectively), and the fact that $\mu$ is traceless, we get $\| \mu \|_2^2 / [d_A(d_A+1)]$.
\end{itemize}
Putting it all together, and using Eq.~\eqref{eq:delta_1} to eliminate $\| \mu \|_2$, we arrive at our main result:
\begin{align}
\Delta^{(k)}
& =  f(k,d_A) \Delta^{(1)} + O((\Delta^{(1)})^2), \label{eq:mainresult_deltak} \\
f(k,d_A) & = \left( \frac{1+d_A}{1+d_A/k} \right)^{1/2}. \label{eq:mainresult_fkd}
\end{align}

\subsection{Scaling of design times \label{sec:designtimes}}

With the result above, Eq.~(\ref{eq:mainresult_deltak}-\ref{eq:mainresult_fkd}), we are now in a position to discuss the scaling of design times in the projected ensemble for our model. 
We begin by observing that, by the definition Eq.~\eqref{eq:distance_measure}, we have 
$\Delta^{(1)} = \sqrt{2^{N_A - S_2(A)} - 1}$, where $S_2(A) = -\log_2 \text{Tr}\rho_A^2$ is the second Renyi entropy of subsystem $A$, measured in bits. 
In chaotic dynamics starting from a disentangled state, we expect the entropy to grow linearly in time~\cite{kim_ballistic_2013}, $S_2(A) \sim v_E t$ (the proportionality constant $v_E$ is known as the entanglement velocity), until near saturation to the maximal value $\log_2(d_A) = N_A$. At that point, achieved when $t \approx t^\ast = N_A/v_E$, we expect a crossover and an exponential convergence to the asymptote: $N_A - S_2(A) \sim 2^{-v_E'(t-t^\ast)}$.

In models described by the statistical mechanics of an entanglement membrane~\cite{nahum_quantum_2017, jonay_coarse-grained_2018, zhou_emergent_2019, zhou_entanglement_2020, li_statistical_2021}, one expects that the two velocities governing the early-time ballistic growth and the late-time exponential convergence should in fact be the same, $v_E\equiv v_E'$, related to the surface tension of the entanglement membrane\footnote{We note however that a ``two-step relaxation process'' (corresponding to two different velocities for the early-time linear growth and the late-time exponential saturation) has been observed in one-dimensional random unitary circuits, depending nontrivially on circuit architecture and boundary conditions~\cite{bensa_fastest_2021, bensa_two-step_2022}.}. 
Our model also displays this behavior in the $q\to\infty$ limit: as shown in Appendix~\ref{app:entropy}, we have
\begin{equation}
    S_2(A) = \left\{
    \begin{aligned}
        & v_E t & (t\ll t^\ast ) \\
        & N_A - \frac{1}{\ln 2} 2^{-v_E (t-t^\ast)} & (t\gg t^\ast )
    \end{aligned}
    \right. ,
    \qquad 
    v_E = 2\log_2(d_{B_1}),
    \qquad 
    t^\ast = N_A/v_E. \label{eq:large_t_ansatz}
\end{equation}
We note that $\log_2(d_{B_1})$ plays the role of $|\partial A|$, the size of the boundary of subsystem $A$ (i.e., the number of qubits with which $A$ interacts directly). In a spatially local system, one indeed expects $v_E \propto |\partial A|$, and $t^\ast \propto |A| / |\partial A| \sim \xi$ if $A$ is a ball of radius $\xi$ in a $d$-dimensional lattice.
Motivated by these considerations, in the following we use $v_E$ for the late-time rate of convergence of $S_2$ towards $N_A$; however this is a slight abuse of notation, and one should be mindful that this identification need not work in general, e.g. in integrable systems, dual-unitary circuits, etc. 

As a consequence of Eq.~\eqref{eq:large_t_ansatz} we have, at late times, 
\begin{equation}
    \Delta^{(1)} = \sqrt{2^{N_A-S_2(A)}-1} \sim 2^{-v_E(t-t^\ast)/2}
    \label{eq:delta1_latetime}
\end{equation}
and thus the thermalization time, defined with respect to an arbitrary threshold $\epsilon \ll 1$:
\begin{equation}
    \Delta^{(1)} < \epsilon
    \quad \iff \quad 
    t > t_1 \equiv \frac{N_A}{v_E} + \frac{2 \log_2(1/\epsilon)}{v_E}.
    \label{eq:t1}
\end{equation}
By using our main result Eq.~(\ref{eq:mainresult_deltak}-\ref{eq:mainresult_fkd}) we can carry out the same reasoning for the $k$-th moment, which gives
\begin{equation}
    \Delta^{(k)} = f(k,d_A) \Delta^{(1)} < \epsilon
    \quad \iff \quad 
    t > t_k \equiv t_1 + \frac{2\log_2 f(k,d_A)}{v_E}.
    \label{eq:tk_separation}
\end{equation}
We have thus obtained the exact design times for this problem, for all $k\geq 1$, in the $q\to \infty$ limit:
\begin{equation}
    t_k = t_1 + \frac{1}{v_E} \log_2 \frac{1+d_A}{1+d_A/k}.
    \label{eq:tk_final}
\end{equation}
This is the second exact calculation of design times in a model of quantum dynamics (after the self-dual kicked Ising model studied in Ref.~\cite{ho_exact_2022}), and the first to show a {\it separation} between distinct design times.
In the following we comment on the scaling of design times with $k$ and $d_A$ in two regimes of interest.

\paragraph{Large $d_A$.}
If $k \ll d_A$, the design times from Eq.~\eqref{eq:tk_final} approximately reduce to 
\begin{equation}
    t_k \approx t_1 + \frac{1}{v_E} \log_2 k,
    \label{eq:log_growth}
\end{equation}
showing a nontrivial growth with $k$. However this growth is only logarithmic in $k$ (in other words, the successive $k$-designs are formed extremely quickly in dynamics, with $k(t) \sim \exp(t)$). 
It is interesting to ask how this compares to other models. Numerical results on chaotic Hamiltonian dynamics at infinite temperature~\cite{cotler_emergent_2021}, while in a limited dynamic range ($1\leq k\leq 4$), are suggestive of a stronger, possibly algebraic growth $t_k \sim k^\alpha$. 
Such a difference in scaling might originate from features of dynamics that are absent in our toy model: geometric locality and energy conservation, both of which could slow down deep thermalization. 
The latter in particular is known to parametrically change the scaling of the thermalization time (due to energy diffusion) and it is interesting to ask how it might affect deep thermalization.

\paragraph{Large $k$.}
The logarithmic growth of $t_k$ in Eq.~\eqref{eq:log_growth} eventually arrests altogether when $k \approx d_A$. This is also the regime in which state designs for different $k$ cease to be fully independent (i.e., for such a $k$, a $k$-design is automatically an approximate $(k+1)$-design~\cite{roberts_chaos_2017}).
At large $k$ we have 
\begin{equation}
    t_\infty = \lim_{k\to\infty} t_k = t_1 + \frac{\log_2(d_A+1)}{v_E} \simeq \frac{2N_A}{v_E} + \frac{2\log_2(1/\epsilon)}{v_E},
    \label{eq:large_k}
\end{equation}
where the approximation holds for $d_A\gg 1$.
The time scale $t_\infty$ governs the emergence of the Haar measure at $A$ (fulfilling the state $k$-design condition for all $k$'s).
Remarkably, we see that 
(i) a uniform distribution of states (to within fixed accuracy $\epsilon$) is generated in finite time in this model, and 
(ii) in the limit $N_A\to\infty$ (taken {\it after} $k\to \infty$) the emergence of the Haar distribution is governed by a renormalized entanglement velocity, $v_\text{Haar} \equiv v_E/2$. 
Appropriately, this takes longer than thermalization; precisely a factor of $2$ longer in this model.


\section{Numerical simulations \label{sec:numerics}}

To test our analytical results, we perform exact numerical simulations of the circuit model. 
We aim to compute the ratio of design distances $\Delta^{(k)} / \Delta^{(1)}$ in order to confirm the relationship in Eq.~(\ref{eq:mainresult_deltak}-\ref{eq:mainresult_fkd}); the scaling of design times follows from the behavior of this ratio.

In order to maximize the accessible environment size, we choose subsystems $A$ and $B_1$ of minimal size, $d_A = d_{B_1} = 2$. We also set $q = 2^{L-2}$ for integer $L$ so that the total Hilbert space corresponds to $L$ qubits.
We initialize a state $\ket{0}^{\otimes L}$ and evolve it under the circuit in Fig.~\ref{fig:schematic}(b), where the gates $U_t$ and $V_t$ are sampled from the Haar measure on $U(4)$ and $U(2^{L-1})$ respectively. After each time step $T = 1, \dots t_\text{max} = 20$ we construct the exact moment operators $\rho^{(k)}$ for $k = 1, \dots k_\text{max} = 7$ and evaluate the distance measures $\Delta^{(k)}(t)$. We repeat this process for many realizations of the Haar-random gates and average the results.
See Appendix~\ref{app:numerics} for further details on the implementation and efficiency of this algorithm.

\begin{figure}
    \centering
    \includegraphics[width=\textwidth]{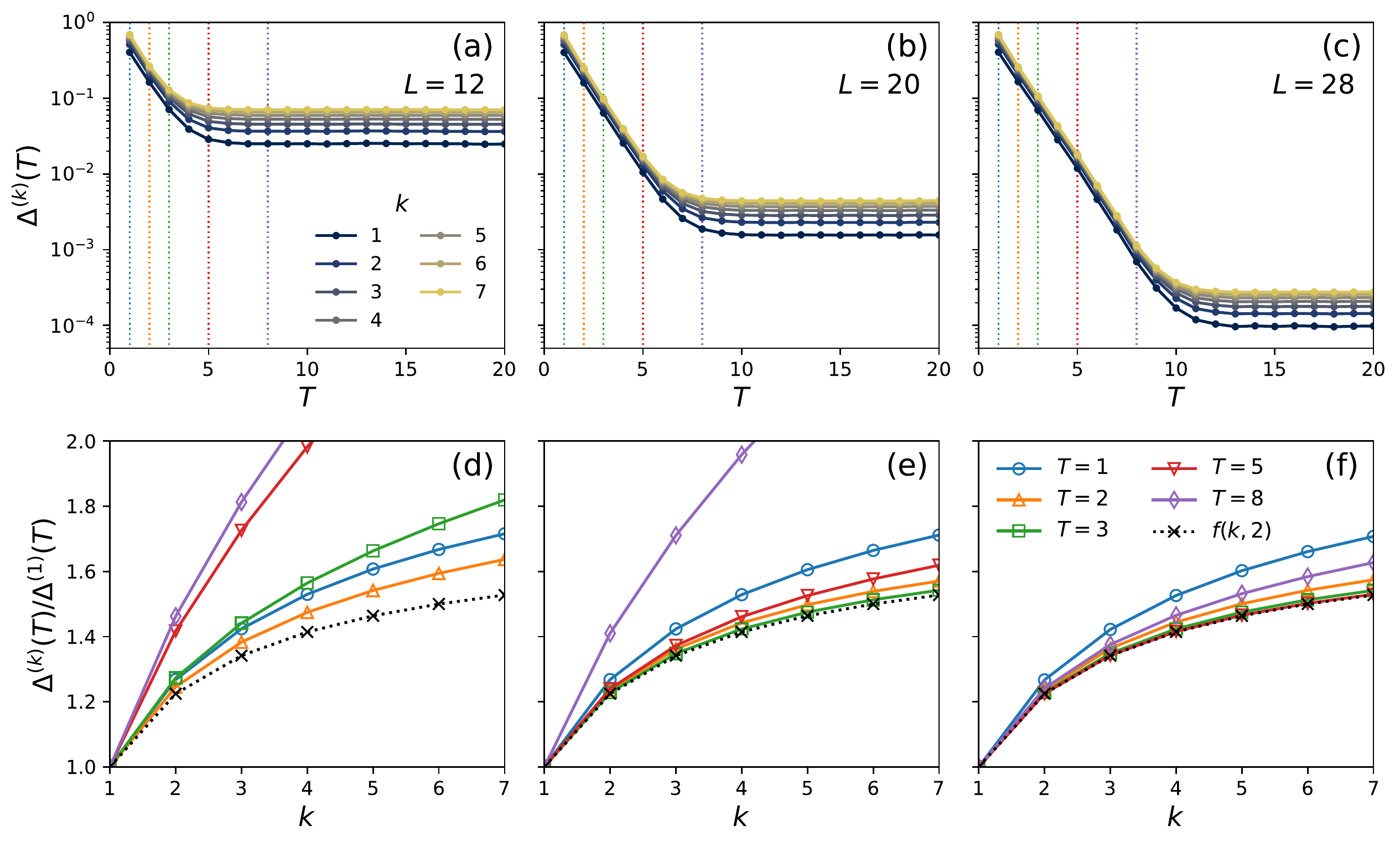}
    \caption{Results of numerical simulations.
    (a-c) $k$-design distance measure $\Delta^{(k)}$ as a function of circuit depth $T$, computed as described in Sec.~\ref{sec:numerics} (see also Appendix~\ref{app:numerics}), averaged over 4000 realizations ($L=12$, 20) and 1000 realizations ($L=28$) of the Haar-random unitary evolution. 
    (d-f) Ratios $\Delta^{(k)} / \Delta^{(1)}$ at various times $T$ (marked by vertical dashed lines of the corresponding color in (a-c)), compared to the analytical prediction $f(k,d_A = 2) = \sqrt{3k/(k+2)}$ (dotted line), Eq.~\eqref{eq:mainresult_fkd}.
    }
    \label{fig:numerics}
\end{figure}

Numerical results for the gate-averaged $\Delta^{(k)}(t)$ are shown in Fig.~\ref{fig:numerics}(a-c), for systems of size up to $L = 28$ (i.e. $q \simeq 6.7\times 10^7$). Focusing first on $k=1$, i.e. regular thermalization, we see an exponential decay followed by saturation to a minimum. 
This final plateau is seen to be a finite-size effect: it corresponds to the finite value one would obtain from a global Haar-random state on $AB$, which is $\sim 1/\sqrt{q}$, as we explain in Appendix~\ref{app:haar}. For $q\to\infty$ the exponential decay would continue indefinitely.

Moving now to $k>1$, we see that the design distances $\Delta^{(k)}$, while qualitatively replicating the behavior of $\Delta^{(1)}$, weakly increase with $k$. The $k$-dependence is visibly slower in the intermediate-time exponential regime and faster in the late-time plateau. In Fig.~\ref{fig:numerics}(d-f) we analyze this $k$-dependence in detail, at different system sizes and times. For times corresponding to the exponential-decay regime, we see an excellent agreement with the analytical prediction 
\begin{equation}
    \frac{\Delta^{(k)}}{ \Delta^{(1)}} \simeq f(k,d_A) = \left( \frac{1+d_A}{1+d_A/k} \right)^{1/2}
\end{equation}
(dashed line). 
The quality of the agreement is limited by two factors:
\begin{itemize}
    \item At early time, $\Delta^{(1)}(T)$ is not yet very small, so the $O((\Delta^{(1)})^2)$ corrections in Eq.~\eqref{eq:mainresult_deltak} are significant. Their effect can be seen clearly e.g. in Fig.~\ref{fig:numerics}(f), with curves for $T=1$, $2$, $3$ showing progressively better agreement.
    \item At later time, finite-$q$ effects become dominant. As soon as the $\Delta^{(k)}(T)$ curves start peeling off from the exponential decay to settle onto the finite-size plateau value $\sim q^{-1/2}$, the agreement with the prediction gets worse. The $k$-dependence crosses over to that of a Haar-random state, which in this case ($d_A = d_{B_1} = 2$) is $\sim \sqrt{k}$, as we show in Appendix~\ref{app:haar}.
\end{itemize}
As a consequence of these two effects, the quality of the agreement with the analytical prediction is non-monotonic: it improves with time at first, and then becomes worse once finite-size effects become substantial.

However, in the $q\to\infty$ limit, finite-size effects would disappear and we would recover the predicted behavior for arbitrarily late times. This is corroborated by the finite-size trends visible in Fig.~\ref{fig:numerics}(d-f): with increasing $q$, the time window of validity of our solution increases, and the agreement itself improves with time within this window. This represents strong evidence in support of our main analytical result, Eq.~(\ref{eq:mainresult_deltak}-\ref{eq:mainresult_fkd}).


\section{Discussion and Conclusion \label{sec:conclusion}}

We have presented a toy model of chaotic quantum dynamics where the process of deep thermalization can be studied exactly in a suitable thermodynamic limit. The simple structure of the model (a subsystem of interest $A$, a ``bottleneck'' $B_1$, and an infinite bath $B_2$) enables this solvability while capturing a basic feature of spatially-local, short-range interacting systems in any dimension: that information may only be exchanged in and out of a subsystem via the mediation of other subsystems. This simple fact has important consequences on the process of thermalization and deep thermalization alike.

Our key result, in Eq.~(\ref{eq:mainresult_deltak}, \ref{eq:mainresult_fkd}), shows how {\it approximate} thermalization, described by a nonzero distance $\Delta^{(1)}$ between the reduced density matrix $\rho_A$ and the infinite-temperature thermal state $\mathbb{I}/d_A$,  translates to a {\it larger} distance $\Delta^{(k)}$ between the higher moments of the projected and Haar ensembles. In other words, the two ensembles become increasingly more distinguishable as one looks at higher moments of their distributions. A nontrivial growth of design times $t_k$ vs $k$ follows: a logarithmic growth for $k\ll d_A$ ($d_A$ being the Hilbert space dimension of $A$), saturating at a finite value $t_\infty = 2t_1$.

This is the second known exact solution for the design times and deep thermalization in models of chaotic dynamics (following the results of Ref.~\cite{ho_exact_2022} on the self-dual kicked Ising chain and its generalization to dual-unitary circuits in one dimension~\cite{claeys_emergent_2022, ippoliti_dynamical_2022}) and it is the first to exhibit distinct time scales $t_k$. This crucially depends on imperfect (or asymptotic) thermalization, which is the generic fate of chaotic many-body systems.
Dual-unitary circuits (with suitable initial states) are exceptional in this regard in that they instead thermalize perfectly in a finite time: one has $\rho_A = \mathbb{I}/d_A$ exactly after a finite time $t_1 = N_A$ in quench dynamics~\cite{piroli_exact_2020}. Thus they evade our present result and show a collapse of design times, $t_k = t_1$ for all $k$. 

Further, it is worth commenting on how our results compare to other bounds on deep thermalization that have been obtained in two settings.
First, Ref.~\cite{wilming_high-temperature_2022} proved that, if one chooses the measurement basis in the bath $B$ Haar-randomly, then the projected ensemble yields an approximate state design with high probability whenever the reduced density matrix is close to thermalization at infinite temperature: $\rho_A\approx \mathbb{I}_A/d_A$. 
More precisely, given a deviation from ideal thermalization $\delta = \frac{1}{2} \| \rho_A - \mathbb{I}_A/d_A\|_1 \leq \frac{1}{2d_A}$, one gets an $\epsilon$-approximate state $k$-design with 
\begin{equation} 
\epsilon \leq 2k\sqrt{d_A\delta} + d_A\delta .
\label{eq:wr_ineq}
\end{equation}
This bound is directly applicable to our model, as the final unitary gate $V_T$ Haar-randomizes the basis on the bath. 
We have found that our model realizes an $\epsilon$-approximate state $k$-design with $\epsilon = f(k,d_A)\Delta^{(1)}$, where $f$ is as in Eq.~\eqref{eq:mainresult_fkd}. 
This implies\footnote{We note that Ref.~\cite{wilming_high-temperature_2022} uses the trace norm, while we use the Frobenius norm; however the norm inequality $\| A \|_2 \leq \| A \|_1$ (for all $A$) gives $\Delta^{(1)} \leq 2\delta \sqrt{d_A}$.} $\epsilon < (2k+1)d_A \delta$; using the assumption that $d_A\delta<1$, and this is seen to be consistent with the bound from Ref.~\cite{wilming_high-temperature_2022}, Eq.~\eqref{eq:wr_ineq}. 
On the other hand, the techniques we used here may be directly adapted to analyze the setting considered in Ref.~\cite{wilming_high-temperature_2022}, 
yielding $\epsilon = \Theta(\delta)$.
This suggests that the bound in Eq.~\eqref{eq:wr_ineq}, which obeys $\epsilon = O(\sqrt{\delta})$ for $\delta \to 0$, may not be tight.  

Secondly, Ref.~\cite{ippoliti_dynamical_2022} derived a bound on $t_\infty$ in one-dimensional unitary circuits: $t_\infty > (v_E / v_p) t_1$, where $v_p$ is an additional velocity scale (the ``purification velocity'').
The derivation of this bound is based on loss of information about measurement outcomes over long distances, and generally only applies in one spatial dimension. 
On the contrary, the exact results about our toy model characterize the effects of imperfect thermalization ($\rho_A \neq \mathbb{I}_A / d_A$, a strictly local condition) on the formation of higher state designs, in a situation that is arguably a best-case scenario for deep thermalization (a large, all-to-all interacting random-matrix bath, as opposed to a structured, geometrically-local one). Thus it is natural to conjecture that the results for this model might instead hold as {\it bounds} for more general models: e.g., we may conjecture that $t_\infty > 2t_1$ in geometrically-local models in any dimension.
As the arguments leading to the two bounds are {\it a priori} independent, one would expect both of them to apply jointly in one-dimensional chaotic unitary circuits, giving $t_\infty > \max \{2, v_E/v_p\} t_1$.
It would be interesting to further investigate the interplay between the two bounds.

Further, it is tempting to speculate that our result for the ratio $\Delta^{(k)} / \Delta^{(1)}$ may be optimal, i.e., that no ensemble consistent with a given density matrix $\rho$ can approximate a $k$-design better than the Scrooge/GAP ensemble; this idea seems supported by the numerical results of Fig.~\ref{fig:numerics}(d-f), where the data never dips below the Scrooge/GAP prediction. Proving this statement is an interesting question for future work.

By a similar reasoning, our results for the design times $t_k$ may also represent bounds in more general settings---e.g., one may conjecture that the growth of $t_k$ with $k$ should be {\it at least} logarithmic (for $k\ll d_A$) in any local, non-perfectly-thermalizing model. This appears consistent with numerical results on Hamiltonian dynamics in spin chains~\cite{cotler_emergent_2021}, where in addition to locality one also has energy diffusion as a possible bottleneck. It would be interesting to understand the impact of conservation laws on deep thermalization, starting e.g.~from quantum circuit models with $U(1)$ charge conservation~\cite{rakovszky_diffusive_2018, khemani_operator_2018, hunter-jones_operator_2018, agrawal_entanglement_2021} or integrable circuit models~\cite{gopalakrishnan_operator_2018, klobas_exact_2021, buca_rule_2021, singh_fredkin_2022}. 
A particularly tractable setting may be fermionic Gaussian dynamics, for which the existence of a limiting ensemble was recently shown~\cite{lucas_generalized_2022}.

\section*{Acknowledgements}
We thank S.~Choi, T.~Rakovszky and V.~Khemani for discussions and for previous collaborations on related topics. M.I.~thanks S.~L.~Sondhi for insightful discussions.
M.~I.~is supported by the Gordon and Betty Moore Foundation's grant GBMF8686.
W.~W.~H.~is supported in part by the Stanford Institute of Theoretical Physics, and by the National University of Singapore (NUS)  start-up grants A-8000599-00-00 and A-8000599-01-00. 
Numerical simulations were performed on Stanford Research Computing Center's Sherlock cluster.

\begin{appendix}


\section{Equivalence between Eq.~\eqref{eqn:limiting_distribution} and the Scrooge/GAP ensemble \label{app:scrooge}}

The Scrooge/GAP ensemble~\cite{jozsa_lower_1994, goldstein_distribution_2006, goldstein_universal_2016} is conventionally defined via {\it ``$\rho$-distortion''} of the Haar measure on $\mathcal{H}_A$: i.e.,   states $\ket{\psi} = \sqrt{\rho_A}\ket{\phi} / \bra{\phi} \rho_A \ket{\phi}$ 
are drawn with weights $d_A \langle \phi| \rho_A | \phi\rangle d\phi$, where $d \phi$ is the Haar measure on $\mathcal{H}_A$.
Here, $\sqrt{\rho_A}$ is the unique positive Hermitian square root of $\rho_A$. 
Our ensemble, Eq.~\eqref{eqn:limiting_distribution}, is instead defined from the Haar measure on a larger Hilbert space $\mathcal{H}_C$ (with $d_C>d_A$) via a rectangular square root of $\rho_A$: the $d_A\times d_C$ matrix $E$, which obeys $EE^\dagger \propto \rho_A$. 
However, the two definitions are equivalent (in the sense that they yield precisely the same ensemble on $\mathcal{H}_A$). This can be seen as follows. From a singular-value decomposition of $E$, we have $E = USPV^\dagger$, where $U$ and $V$ are unitary matrices on $\mathcal{H}_A$ and $\mathcal{H}_C$ respectively, $S$ is a positive diagonal $d_A\times d_A$ matrix of singular values, and $P$ is a rectangular $d_A\times d_C$ matrix that projects onto the first $d_A$ entries (i.e., $P_{ij} = \delta_{ij}$ if $j\leq d_A$, 0 otherwise). 
From $\rho_A\propto EE^\dagger = US^2U^\dagger$ we have $E \propto \sqrt{\rho_A} U P V^\dagger$; further noting that $UP = P(U\otimes \mathbb{I})$, where the identity operator has size $d_C/d_A$, and absorbing $(U\otimes \mathbb{I})V^\dagger$ in the Haar measure on $\mathcal{H}_C$, we see that in the ensemble of Eq.~\eqref{eqn:limiting_distribution} we may replace $E$ by $\sqrt{\rho_A} P$. Finally, we see that $P \ket{\phi} / \| P\ket{\phi}\|$ is a Haar-random vector of size $d_A$ (i.e., a projector acting on a Haar-random state produces another Haar-random state of the appropriate size, up to normalization). We can thus replace the Haar measure ${\rm d}\phi$ on $C$ with the analogous measure on $A$, obtaining the usual $\rho$-distortion construction of the Scrooge/GAP ensemble. 
 

\section{Thermalization of subsystem $A$ \label{app:entropy}}

Here we derive the exact form of the purity of subsystem $A$ as a function of time $T$, in the limit $q\to\infty$, averaged over Haar-random gates $U_t$. 

\begin{figure}
    \centering
    \includegraphics[width=0.7\textwidth]{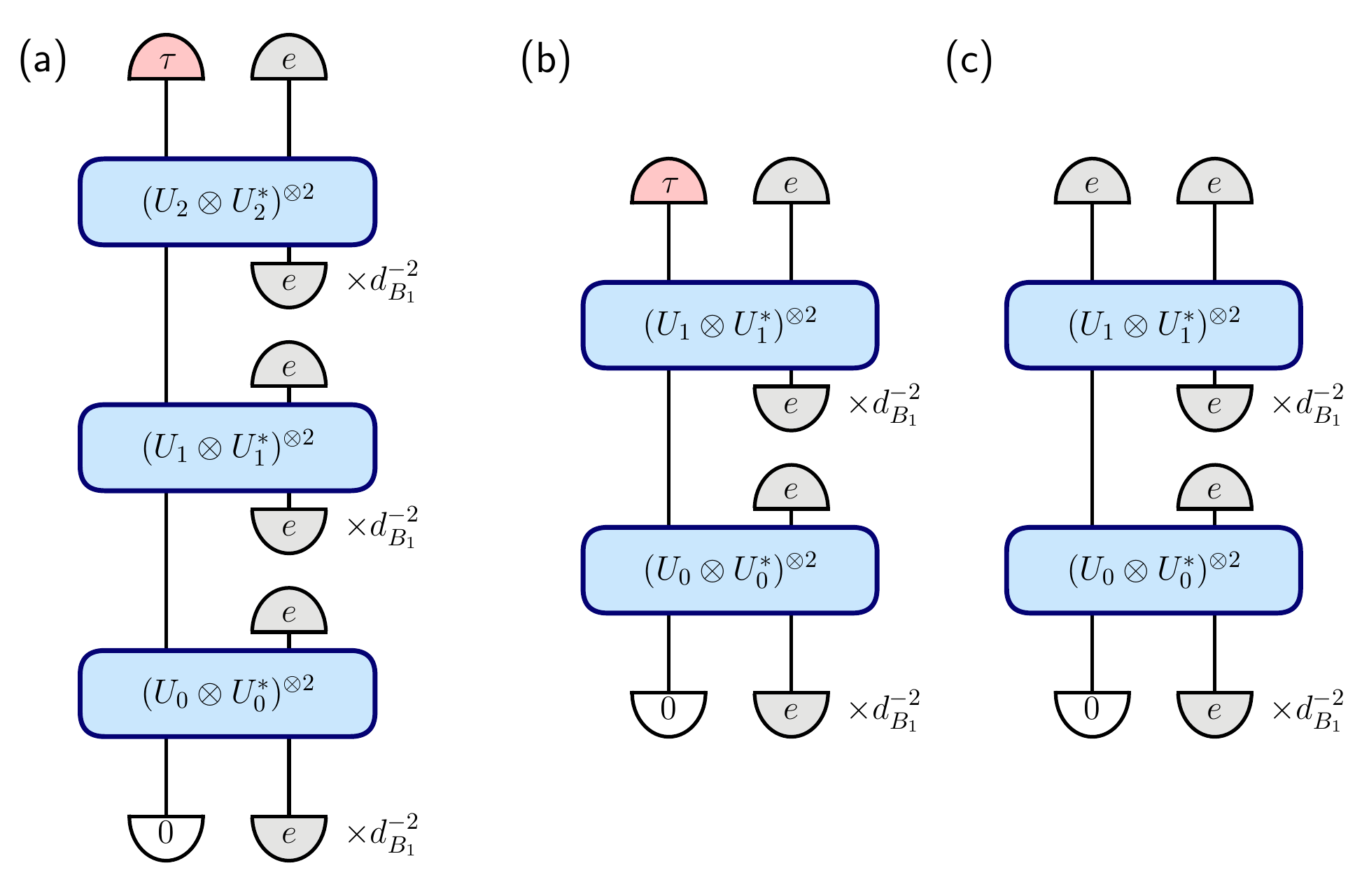}
    \caption{Purity of subsystem $A$. 
    (a) Tensor network contraction expressing ${\rm Tr}(\rho_A^2)$ after evolution for time $T = 3$. The $|\cdots)$ symbols denote operators on two replicas of the appropriate Hilbert space, with $|0) = \ketbra{0}{0}^{\otimes 2}$ (initial state on $A$), and $|e)$ and $|\tau)$ are replica permutations. Note that $d_{B_1}^{-2} |e)$ corresponds to two replicas of the completely-mixed state on $B_1$, $(\mathbb{I}_{B_1}/d_{B_1})^{\otimes 2}$. 
    Upon averaging $U_{T-1}$ over the Haar measure one obtains two terms: (b) the same diagram as in (a), but with $T\mapsto T-1$; (c) a diagram with a uniform $|e)$ boundary condition at the final time, which simply computes the trace of the state and is trivial due to trace-preservation. This gives the recursion in Eq.~\eqref{eq:purity_recursion}.
    }
    \label{fig:therm}
\end{figure}

Starting from the reduced density matrix $\rho_A$ as written in Eq.~\eqref{eq:rdm}, we have
\begin{equation}
    P_A(T) \equiv {\rm Tr} \rho_A^2(T)
    = d_{B_1}^{-2T} {\rm Tr} (EE^\dagger)^2.
    \label{eq:purity_app}
\end{equation}
This is represented diagrammatically in Fig.~\ref{fig:therm}(a), where the boundary conditions are permutations of the two replicas, $e,\tau \in S_2$: the identity $e$ before and after each gate on $B_1$, and the transposition $\tau$ at the final time on $A$. We note that $d_{B_1}^{-2} |e)(e|_{B_1}$ is a super-operator implementing the fully depolarizing (or erasure) channel on each replica of $B_1$---i.e., it is the partial trace on $B_1$ followed by introducing the fully-mixed state $\mathbb{I}_{B_1} / d_{B_1}$. This Markovian evolution is a consequence of taking the $q\to\infty$ limit on $B_2$, which introduces irreversibility in the local dynamics of $A$ and $B_1$. 

We compute the purity, Eq.~\eqref{eq:purity_app}, averaged over the choice of gates $\{U_t: t=0,\dots T-1\}$ which make up $E$. Each $U_t$ is sampled independently and identically form the Haar measure.
We compute the average iteratively, starting from the last gate, $U_{T-1}$. Acting on the operator $|\hat \tau )_A\otimes |\hat e)_{B_1}$ above it, the Haar-averaged gate yields
\begin{equation}
    [\mathbb{E} (U\otimes U^\ast)^{\otimes 2}] |\hat \tau)_A \otimes |\hat e)_{B_1}
    = \frac{(d_A^2-1)d_{B_1}}{d_A^2 d_{B_1}^2-1} |\hat \tau)_A \otimes |\hat \tau)_{B_1}
    + \frac{d_A(d_{B_1}^2-1)}{d_A^2 d_{B_1}^2-1} |\hat e)_A \otimes |\hat e)_{B_1}. \label{eq:gateavg}
\end{equation}
This result admits a simple interpretation. At the final time $T$ we have a ``domain wall'' boundary condition, with permutation $\tau$ in $A$ and $e$ in $B_1$. This configuration is evolved backward in time and gives two possible configurations: one where the whole system is polarized along $e$, and one where it is polarized along $\tau$. 
This structure of the Haar average for second-moment quantities was used previously to obtain exact results on entanglement growth, by mapping to a simple random walk for an extended $1+1$-dimensional quantum circuit~\cite{nahum_quantum_2017, von_keyserlingk_operator_2018, ippoliti_fractal_2022}.

Moving down along the tensor network, we must contract the result of Eq.~\eqref{eq:gateavg} with $d_{B_1}^{-2} |\hat e)(\hat e|_{B_1}$ (i.e. the erasure channel on $B_1$), obtaining
\begin{equation}
    \frac{1}{d_{B_1}^2} |\hat e)(\hat e|_{B_1} [\mathbb{E} (U\otimes U^\ast)^{\otimes 2}] |\hat \tau)_A \otimes |\hat e)_{B_1}
    = \frac{d_A^2-1}{d_A^2 d_{B_1}^2-1} |\hat \tau)_A \otimes |\hat e)_{B_1}
    + \frac{d_A(d_{B_1}^2-1)}{d_A^2 d_{B_1}^2-1} |\hat e)_A \otimes |\hat e)_{B_1}
\end{equation}
(we use the fact that $(\hat e|\hat e)_{B_1} = d_{B_1}^2$ and $(\hat e|\hat \tau)_{B_1} = d_{B_1}$). 
The first term, shown in Fig.~\ref{fig:therm}(b), is identical to the initial boundary condition (a domain wall between $\tau$ and $e$) up to a prefactor; 
the second term, shown in Fig.~\ref{fig:therm}(c), is a uniform $|e)_{AB_1}$ state, which is invariant under the remaining backward evolution\footnote{This follows from trace preservation, and thus holds under any quantum channels, including in particular the unitary and depolarizing channels considered in this calculation.}. 
This is enough to derive the following recursion:
\begin{equation}
    \mathbb{E} P_A (T) 
    = \frac{(d_A^2-1) \mathbb{E} P_A (T-1) + d_A(d_{B_1}^2-1)}{d_A^2 d_{B_1}^2-1} . \label{eq:purity_recursion}
\end{equation}
The recursion is solved exactly (with initial condition $P_A(0) = 1$) by 
\begin{equation}
    \mathbb{E} P_A(T) = \frac{1}{d_A} + \left( 1-\frac{1}{d_A}\right) \left[ \frac{d_A^2-1}{d_A^2 d_{B_1}^2-1} \right]^T
    \label{eq:recursion_solution}
\end{equation}
This obeys several sanity checks: $d_A^{-1} < P_A(T) \leq 1$ (corresponding to a valid purity), $P_A(\infty) = d_A^{-1}$ (corresponding to the completely mixed state), and $P_A(T)\equiv 1$ for $d_{B_1} = 1$ (trivial subsystem $B_1$, so that $A$ is in fact isolated and remains pure).

For a large subsystem $d_A \gg 1$, Eq.~\eqref{eq:recursion_solution} reduces to $P_A(T) \simeq d_A^{-1} + d_B^{-2T}$.
This has two limiting behaviors for the annealed average of the second Renyi entropy:
\begin{equation}
    S_2(A,T) = -\log_2\mathbb{E} P_A(T) =
    \left\{ 
    \begin{aligned}
        & 2\log_2(d_{B_1})T & \text{ if } d_{B_1}^{2T} \ll d_A \\
        & N_A - \frac{1}{\ln 2} d_A d_{B_1}^{-2T} & \text{ if } d_{B_1}^{2T} \gg d_A 
    \end{aligned}
    \right.
\end{equation}
which yields Eq.~\eqref{eq:large_t_ansatz}, proving our claim that the early-time slope of $S_2(A,T)$ coincides with the time constant of the exponential approach to the asymptote $S_2(A) = N_A$.

This result is also expected to hold more generally in extended chaotic systems where entanglement is described by a `minimal membrane' picture~\cite{nahum_quantum_2017, zhou_emergent_2019, zhou_entanglement_2020, jonay_coarse-grained_2018, li_statistical_2021}. 
There, the entropy is given by the free energy of a classical statistical system with a partition function $\mathcal{Z} = \sum_\gamma e^{-E(\gamma)}$. 
Here $\gamma$ labels membranes that partition the spacetime volume into regions whose boundaries are $A$ and its complement, and $E(\gamma)$ is a suitable cost functional akin to a surface tension.
In a local quantum circuit of depth $T$ in arbitrary spatial dimension $d$, for a contiguous region $A$, there are two distinct kinds of membranes to consider: temporal ones, $E(\gamma) \sim |\partial A|T$, and spatial ones, $E(\gamma) \sim N_A$. The former are dominant at small $T$, giving ballistic growth of entanglement, motivating the definition of an entanglement velocity $v_E$ via $E(\gamma) = v_E T$. The latter become dominant at large $T$, giving saturation to a volume-law entangled state.
In general we have $S \simeq - \log_2( 2^{-N_A} + 2^{-v_E T})$, which, expanded at large $T$, gives
\begin{equation}
    S \simeq N_A -\log_2(1 + 2^{N_A - v_ET})
    \simeq N_A - \frac{1}{\ln 2} 2^{-v_E(T-t^\ast)}, 
    \quad
    t^\ast = N_A/v_E.
\end{equation}

Note however that exceptions to the above behavior have been recently discovered in finite one-dimensional random-unitary circuits with certain architectures and boundary conditions~\cite{bensa_fastest_2021, bensa_two-step_2022}.


\section{Details on computational method \label{app:numerics}}

Here we present some details on the numerical method used to produce the data in Fig.~\ref{fig:numerics}.
As explained in Sec.~\ref{sec:numerics}, we use $d_A = d_{B_1} = 2$ and $q = 2^{L-2}$, and we begin with the all-zero state $\ket{\Psi(0)} = \ket{0}^{\otimes L}$. There are two computational tasks to discuss: how to implement time evolution, and how to build the moment operator $\rho^{(k)}$ from a given state.

\paragraph{Time evolution.}
The action of gates $U_t \in U(4)$ is implemented via dense matrix-matrix multiplication ($\Psi$ being viewed as a $4\times q$ matrix) which takes time $O(q)$. Doing the same thing for $V_t$ would take time $O(q^2)$; however this is not necessary. Instead we use the following algorithm:
\begin{enumerate}
    \item decompose
    \begin{equation}
        \ket{\Psi(t)} = c_0 \ket{0}_A \ket{\chi_0}_B + c_1 \ket{1}_A \ket{\chi_1}_B,
    \end{equation}
    where $\ket{\chi_0}$, $\ket{\chi_1}$ are normalized; 
    \item compute the overlap $\braket{\chi_0}{\chi_1} \equiv \cos\theta e^{i\varphi}$;
    \item generate two orthonormal Haar-random vectors $\ket{\eta_0}$, $\ket{\eta_1}$;
    \item update $\ket{\Psi(t)} \mapsto c_0 \ket{0}_A \ket{\eta_0}_B + c_1 \ket{1}_A (\cos\theta e^{i\varphi} \ket{\eta_0} + \sin\theta \ket{\eta_1})_B$.
\end{enumerate}
This method only deals with the action of $V_t$ on the subspace spanned by $\ket{\chi_0}$ and $\ket{\chi_1}$, which is all that is needed to evolve $\ket{\Psi}$, and requires time $O(q)$.

\paragraph{Moment operator.}
For a fixed value of $k$, $\rho^{(k)}$ lives in a replicated Hilbert space $\mathcal{H}_A^{\otimes k}$ of dimension $d_A^k$; however it is supported only in the symmetric sector, of dimension $\binom{k+d_A-1}{k}$. In the case of interest, $d_A=2$, this dimension reduces to $k+1$: a basis is given by symmetrized versions of $\ket{0^{\otimes l} 1^{\otimes k-l}}$ for $l=0, \dots k$. 
All the information in $\rho^{(k)}$ is thus obtained from the matrix elements
\begin{equation}
    \bra{0^{\otimes l_1} 1^{\otimes k-l_1}} \rho^{(k)} \ket{0^{\otimes l_2} 1^{\otimes k-l_2}}
    = \sum_z p(z) \braket{0}{\psi_z}^{l_1} \braket{1}{\psi_z}^{k-l_1} \braket{\psi_z}{0}^{l_2} \braket{\psi_z}{1}^{k-l_2}
\end{equation}
for $l_1\leq l_2$ (the remaining terms are given by Hermiticity). For fixed $l_1$, $l_2$ the sum is computed in time $O(q)$ and thus the overall time cost to get $\rho^{(k)}$ is $O(qk^2)$. 
Iterating this process for all $k$ up to $k_\text{max}$ costs $O(qk_\text{max}^3)$. 

Note that for general $d_A$ one has $\binom{d_A-1+k}{k} \sim k^{d_A-1}$ (at large $k$), giving total cost $O(qk_\text{max}^{2d_A-1})$; thus, in order to analyze the largest possible range in $k$, we set $d_A = 2$.

\section{$\Delta^{(k)}$ for Haar-random states \label{app:haar}}
Here we derive the form of the design distance measures $\Delta^{(k)}$ for Haar-random states on a bipartite system $AB$ and show that it matches the post-saturation behavior observed in our numerical simulations, Fig.~\ref{fig:numerics}.
We employ the same replica trick as in Sec.~\ref{sec:env} and compute the Haar-average of the pseudo-frame potential 
\begin{equation}
    F^{(n,k)} = \sum_{z_1, z_2=1}^{d_B} 
    \braket{\tilde \psi_{z_1} }{\tilde \psi_{z_1}}^n 
    |\braket{\tilde \psi_{z_1}}{\tilde \psi_{z_2}}|^{2k} 
    \braket{\tilde \psi_{z_2} }{\tilde \psi_{z_2}}^n 
    = {\rm Tr}(\mathcal{Q}_B \ketbra{\Psi}{\Psi}_{AB}^{\otimes R} )
\end{equation}
where $\ket{\Psi}_{AB}$ is the state of $AB$ (to be averaged over the Haar measure) and $\ket{\tilde{\psi}_z}_A = {}_B\!\braket{z}{\Psi}_{AB}$ is the unnormalized post-measurement state of $A$. The operator $\mathcal{Q}_B$ is as in Eq.~\eqref{eq:Q} and $R = 2(n+k)$ is the total number of replicas.
On average we have
\begin{equation}
    \mathbb{E} F^{(n,k)} 
    = {\rm Tr}(\mathcal{Q}_B \rho_{H,AB}^{(R)})
    = \frac{(d_A d_B-1)!}{(R+d_Ad_B-1)!} \sum_{\sigma\in S_R} {\rm Tr}(\mathcal{Q}_B \hat{\sigma}_{AB}).
\end{equation}
Next, we have $(\mathcal{Q}|\hat{\sigma})_B = d_B + d_B(d_B-1) \delta_{\sigma, \tau \pi_1 \pi_2}$, where $\tau$ and $\pi_{1,2}$ are as in Fig.~\ref{fig:env_diagrams}(e) and the two terms come from expanding $\mathcal{Q}$ into $z_1=z_2$ and $z_1\neq z_2$ terms. 
Following the same manipulations as in Sec.~\ref{sec:env}, one gets
\begin{equation}
    \sum_{\sigma\in S_R} {\rm Tr}(\mathcal {Q}_B \hat{\sigma}_{AB})
    = d_B \frac{(R+d_A-1)!}{(d_A-1)!} + d_B (d_B-1) \left[\frac{(R/2+d_A-1)!}{(d_A-1)!}\right]^2 F_H^{(k)}
\end{equation}
and thus, taking $n = 1-k$ (i.e. $R = 2$),
\begin{equation}
    \mathbb{E} F^{(k)} = \frac{d_A+1 + d_A(d_B-1) F_H^{(k)}}{d_Ad_B-1}
\end{equation}
Finally, we obtain
\begin{equation}
    \mathbb{E} (\Delta^{(k)})^2 
    = \frac{\mathbb{E}F^{(k)}}{F_H^{(k)}} -1 
    = \frac{d_A+1}{d_Ad_B-1}\left[ \binom{k+d_A-1}{k} -1 \right].
\end{equation}
Incidentally, we note that this identity may serve as an alternative approach to derive the results of Ref.~\cite{cotler_emergent_2021} on obtaining an $\epsilon$-approximate $k$-design from the projected ensemble on a Haar-random state with a given probability.

Thus far we have made no assumptions on $d_{A,B}$. Next, we plug in the values used in our numerical simulations, see Sec.~\ref{sec:numerics}: $d_A = 2$, $d_B = 2q$. We find
\begin{equation}
    (\Delta^{(k)})_\text{r.m.s.} 
    = [\mathbb{E}(\Delta^{(k)})^2]^{1/2}
    = \left( \frac{3k}{4q+1} \right)^{1/2}.
\end{equation}
This shows at once that we have a finite-size ``floor'' $\Delta^{(1)}\sim q^{-1/2}$, attained when the dynamics has successfully Haar-randomized the entire system $AB$, and that in this regime the $k$-dependence is $\Delta^{(k)} = \sqrt{k}\Delta^{(1)}$.
Notably the ratio $\Delta^{(k)} / \Delta^{(1)}$ is unbounded, in contrast with our result Eq.~\eqref{eq:mainresult_deltak} which for the same value of $d_A$ gives $\Delta^{(k)} = \sqrt{3k/(k+2)}\Delta^{(1)} \leq \sqrt{3} \Delta^{(1)}$.
This accounts for the distinct behaviors visible in Fig.~\ref{fig:numerics}(a-c) in the two regimes.

\end{appendix}

\bibliographystyle{quantum}
\bibliography{designs}

\begin{thebibliography}{10}

\bibitem{altman_quantum_2021}
Ehud Altman, Kenneth~R. Brown, Giuseppe Carleo, Lincoln~D. Carr, Eugene Demler,
  Cheng Chin, et~al.
\newblock ``Quantum {Simulators}: {Architectures} and {Opportunities}''.
\newblock \href{https://dx.doi.org/10.1103/PRXQuantum.2.017003}{PRX Quantum
  {\bf 2}, 017003}~(2021).

\bibitem{bernien_probing_2017}
Hannes Bernien, Sylvain Schwartz, Alexander Keesling, Harry Levine, Ahmed
  Omran, Hannes Pichler, Soonwon Choi, Alexander~S. Zibrov, Manuel Endres,
  Markus Greiner, Vladan Vuleti{\'c}, and Mikhail~D. Lukin.
\newblock ``Probing many-body dynamics on a 51-atom quantum simulator''.
\newblock \href{https://dx.doi.org/10.1038/nature24622}{Nature {\bf 551},
  579--584}~(2017).

\bibitem{choi_observation_2017}
Soonwon Choi, Joonhee Choi, Renate Landig, Georg Kucsko, Hengyun Zhou, Junichi
  Isoya, Fedor Jelezko, Shinobu Onoda, Hitoshi Sumiya, Vedika Khemani, Curt von
  Keyserlingk, Norman~Y. Yao, Eugene Demler, and Mikhail~D. Lukin.
\newblock ``Observation of discrete time-crystalline order in a disordered
  dipolar many-body system''.
\newblock \href{https://dx.doi.org/10.1038/nature21426}{Nature {\bf 543},
  221--225}~(2017).

\bibitem{mi_time-crystalline_2022}
Xiao Mi, Matteo Ippoliti, Chris Quintana, Ami Greene, Zijun Chen, Jonathan
  Gross, et~al.
\newblock ``Time-crystalline eigenstate order on a quantum processor''.
\newblock \href{https://dx.doi.org/10.1038/s41586-021-04257-w}{Nature {\bf
  601}, 531--536}~(2022).

\bibitem{randall_many-bodylocalized_2021}
J.~Randall, C.~E. Bradley, F.~V. van~der Gronden, A.~Galicia, M.~H. Abobeih,
  M.~Markham, D.~J. Twitchen, F.~Machado, N.~Y. Yao, and T.~H. Taminiau.
\newblock ``Many-body{\textendash}localized discrete time crystal with a
  programmable spin-based quantum simulator''.
\newblock \href{https://dx.doi.org/10.1126/science.abk0603}{Science {\bf 374},
  1474--1478}~(2021).

\bibitem{dumitrescu_dynamical_2022}
Philipp~T. Dumitrescu, Justin~G. Bohnet, John~P. Gaebler, Aaron Hankin, David
  Hayes, Ajesh Kumar, Brian Neyenhuis, Romain Vasseur, and Andrew~C. Potter.
\newblock ``Dynamical topological phase realized in a trapped-ion quantum
  simulator''.
\newblock \href{https://dx.doi.org/10.1038/s41586-022-04853-4}{Nature {\bf
  607}, 463--467}~(2022).

\bibitem{skinner_measurement-induced_2019}
Brian Skinner, Jonathan Ruhman, and Adam Nahum.
\newblock ``Measurement-{Induced} {Phase} {Transitions} in the {Dynamics} of
  {Entanglement}''.
\newblock \href{https://dx.doi.org/10.1103/PhysRevX.9.031009}{Physical Review X
  {\bf 9}, 031009}~(2019).

\bibitem{li_quantum_2018}
Yaodong Li, Xiao Chen, and Matthew P.~A. Fisher.
\newblock ``Quantum {Zeno} effect and the many-body entanglement transition''.
\newblock \href{https://dx.doi.org/10.1103/PhysRevB.98.205136}{Physical Review
  B {\bf 98}, 205136}~(2018).

\bibitem{choi_quantum_2020}
Soonwon Choi, Yimu Bao, Xiao-Liang Qi, and Ehud Altman.
\newblock ``Quantum {Error} {Correction} in {Scrambling} {Dynamics} and
  {Measurement}-{Induced} {Phase} {Transition}''.
\newblock \href{https://dx.doi.org/10.1103/PhysRevLett.125.030505}{Physical
  Review Letters {\bf 125}, 030505}~(2020).

\bibitem{gullans_dynamical_2020}
Michael~J. Gullans and David~A. Huse.
\newblock ``Dynamical {Purification} {Phase} {Transition} {Induced} by
  {Quantum} {Measurements}''.
\newblock \href{https://dx.doi.org/10.1103/PhysRevX.10.041020}{Physical Review
  X {\bf 10}, 041020}~(2020).

\bibitem{ippoliti_entanglement_2021}
Matteo Ippoliti, Michael~J. Gullans, Sarang Gopalakrishnan, David~A. Huse, and
  Vedika Khemani.
\newblock ``Entanglement {Phase} {Transitions} in {Measurement}-{Only}
  {Dynamics}''.
\newblock \href{https://dx.doi.org/10.1103/PhysRevX.11.011030}{Physical Review
  X {\bf 11}, 011030}~(2021).

\bibitem{potter_entanglement_2021}
Andrew~C. Potter and Romain Vasseur.
\newblock ``Entanglement dynamics in hybrid quantum circuits''~(2021).
\newblock  \href{http://arxiv.org/abs/2111.08018}{arXiv:2111.08018}.

\bibitem{fisher_random_2022}
Matthew P.~A. Fisher, Vedika Khemani, Adam Nahum, and Sagar Vijay.
\newblock ``Random {Quantum} {Circuits}''~(2022).
\newblock  \href{http://arxiv.org/abs/2207.14280}{arXiv:2207.14280}.

\bibitem{noel_measurement-induced_2022}
Crystal Noel, Pradeep Niroula, Daiwei Zhu, Andrew Risinger, Laird Egan,
  Debopriyo Biswas, Marko Cetina, Alexey~V. Gorshkov, Michael~J. Gullans,
  David~A. Huse, and Christopher Monroe.
\newblock ``Measurement-induced quantum phases realized in a trapped-ion
  quantum computer''.
\newblock \href{https://dx.doi.org/10.1038/s41567-022-01619-7}{Nature Physics
  {\bf 18}, 760--764}~(2022).

\bibitem{koh_experimental_2022}
Jin~Ming Koh, Shi-Ning Sun, Mario Motta, and Austin~J. Minnich.
\newblock ``Experimental {Realization} of a {Measurement}-{Induced}
  {Entanglement} {Phase} {Transition} on a {Superconducting} {Quantum}
  {Processor}''~(2022).
\newblock  \href{http://arxiv.org/abs/2203.04338}{arXiv:2203.04338}.

\bibitem{aaronson_shadow_2018}
Scott Aaronson.
\newblock ``Shadow {Tomography} of {Quantum} {States}''~(2018).
\newblock  \href{http://arxiv.org/abs/1711.01053}{arXiv:1711.01053}.

\bibitem{huang_predicting_2020}
Hsin-Yuan Huang, Richard Kueng, and John Preskill.
\newblock ``Predicting many properties of a quantum system from very few
  measurements''.
\newblock \href{https://dx.doi.org/10.1038/s41567-020-0932-7}{Nature Physics
  {\bf 16}, 1050--1057}~(2020).

\bibitem{elben_randomized_2022}
Andreas Elben, Steven~T. Flammia, Hsin-Yuan Huang, Richard Kueng, John
  Preskill, Beno{\^i}t Vermersch, and Peter Zoller.
\newblock ``The randomized measurement toolbox''~(2022).
\newblock  \href{http://arxiv.org/abs/2203.11374}{arXiv:2203.11374}.

\bibitem{arute_quantum_2019}
Frank Arute, Kunal Arya, Ryan Babbush, Dave Bacon, Joseph~C. Bardin, Rami
  Barends, et~al.
\newblock ``Quantum supremacy using a programmable superconducting processor''.
\newblock \href{https://dx.doi.org/10.1038/s41586-019-1666-5}{Nature {\bf 574},
  505--510}~(2019).

\bibitem{wu_strong_2021}
Yulin Wu, Wan-Su Bao, Sirui Cao, Fusheng Chen, Ming-Cheng Chen, Xiawei Chen,
  et~al.
\newblock ``Strong {Quantum} {Computational} {Advantage} {Using} a
  {Superconducting} {Quantum} {Processor}''.
\newblock \href{https://dx.doi.org/10.1103/PhysRevLett.127.180501}{Physical
  Review Letters {\bf 127}, 180501}~(2021).

\bibitem{aaronson_complexity-theoretic_2017}
Scott Aaronson and Lijie Chen.
\newblock ``Complexity-theoretic foundations of quantum supremacy
  experiments''.
\newblock In Proceedings of the 32nd {Computational} {Complexity} {Conference}.
\newblock \href{https://dx.doi.org/10.5555/3135595.3135617}{Pages 1--67}.
\newblock {CCC} '17Dagstuhl, DEU~(2017).

\bibitem{zhou_what_2020}
Yiqing Zhou, E.~Miles Stoudenmire, and Xavier Waintal.
\newblock ``What {Limits} the {Simulation} of {Quantum} {Computers}?''.
\newblock \href{https://dx.doi.org/10.1103/PhysRevX.10.041038}{Physical Review
  X {\bf 10}, 041038}~(2020).

\bibitem{gao_limitations_2021}
Xun Gao, Marcin Kalinowski, Chi-Ning Chou, Mikhail~D. Lukin, Boaz Barak, and
  Soonwon Choi.
\newblock ``Limitations of {Linear} {Cross}-{Entropy} as a {Measure} for
  {Quantum} {Advantage}''~(2021).
\newblock  \href{http://arxiv.org/abs/2112.01657}{arXiv:2112.01657}.

\bibitem{deshpande_tight_2021}
Abhinav Deshpande, Pradeep Niroula, Oles Shtanko, Alexey~V. Gorshkov, Bill
  Fefferman, and Michael~J. Gullans.
\newblock ``Tight bounds on the convergence of noisy random circuits to the
  uniform distribution''~(2021).
\newblock  \href{http://arxiv.org/abs/2112.00716}{arXiv:2112.00716}.

\bibitem{dalzell_random_2021}
Alexander~M. Dalzell, Nicholas Hunter-Jones, and Fernando G. S.~L. Brand{\~a}o.
\newblock ``Random quantum circuits transform local noise into global white
  noise''~(2021).
\newblock  \href{http://arxiv.org/abs/2111.14907}{arXiv:2111.14907}.

\bibitem{choi_emergent_2021}
Joonhee Choi, Adam~L. Shaw, Ivaylo~S. Madjarov, Xin Xie, Jacob~P. Covey,
  Jordan~S. Cotler, Daniel~K. Mark, Hsin-Yuan Huang, Anant Kale, Hannes
  Pichler, Fernando G. S.~L. Brand{\~a}o, Soonwon Choi, and Manuel Endres.
\newblock ``Emergent {Randomness} and {Benchmarking} from {Many}-{Body}
  {Quantum} {Chaos}''~(2021).
\newblock  \href{http://arxiv.org/abs/2103.03535}{arXiv:2103.03535}.

\bibitem{mark_benchmarking_2022}
Daniel~K. Mark, Joonhee Choi, Adam~L. Shaw, Manuel Endres, and Soonwon Choi.
\newblock ``Benchmarking {Quantum} {Simulators} using {Quantum}
  {Chaos}''~(2022).
\newblock  \href{http://arxiv.org/abs/2205.12211}{arXiv:2205.12211}.

\bibitem{srednicki_chaos_1994}
Mark Srednicki.
\newblock ``Chaos and quantum thermalization''.
\newblock \href{https://dx.doi.org/10.1103/PhysRevE.50.888}{Physical Review E
  {\bf 50}, 888--901}~(1994).

\bibitem{rigol_thermalization_2008}
Marcos Rigol, Vanja Dunjko, and Maxim Olshanii.
\newblock ``Thermalization and its mechanism for generic isolated quantum
  systems''.
\newblock \href{https://dx.doi.org/10.1038/nature06838}{Nature {\bf 452},
  854--858}~(2008).

\bibitem{abanin_colloquium_2019}
Dmitry~A. Abanin, Ehud Altman, Immanuel Bloch, and Maksym Serbyn.
\newblock ``Colloquium: {Many}-body localization, thermalization, and
  entanglement''.
\newblock \href{https://dx.doi.org/10.1103/RevModPhys.91.021001}{Reviews of
  Modern Physics {\bf 91}, 021001}~(2019).

\bibitem{alhambra_quantum_2022}
{\'A}lvaro~M. Alhambra.
\newblock ``Quantum many-body systems in thermal equilibrium''~(2022).
\newblock  \href{http://arxiv.org/abs/2204.08349}{arXiv:2204.08349}.

\bibitem{cotler_emergent_2021}
Jordan~S. Cotler, Daniel~K. Mark, Hsin-Yuan Huang, Felipe Hernandez, Joonhee
  Choi, Adam~L. Shaw, Manuel Endres, and Soonwon Choi.
\newblock ``Emergent quantum state designs from individual many-body
  wavefunctions''~(2021).
\newblock  \href{http://arxiv.org/abs/2103.03536}{arXiv:2103.03536}.

\bibitem{ho_exact_2022}
Wen~Wei Ho and Soonwon Choi.
\newblock ``Exact {Emergent} {Quantum} {State} {Designs} from {Quantum}
  {Chaotic} {Dynamics}''.
\newblock \href{https://dx.doi.org/10.1103/PhysRevLett.128.060601}{Physical
  Review Letters {\bf 128}, 060601}~(2022).

\bibitem{ippoliti_dynamical_2022}
Matteo Ippoliti and Wen~Wei Ho.
\newblock ``Dynamical purification and the emergence of quantum state designs
  from the projected ensemble''~(2022).
\newblock  \href{http://arxiv.org/abs/2204.13657}{arXiv:2204.13657}.

\bibitem{ambainis_quantum_2007}
Andris Ambainis and Joseph Emerson.
\newblock ``Quantum t-designs: t-wise {Independence} in the {Quantum}
  {World}''.
\newblock In Twenty-{Second} {Annual} {IEEE} {Conference} on {Computational}
  {Complexity} ({CCC}'07).
\newblock \href{https://dx.doi.org/10.1109/CCC.2007.26}{Pages 129--140}.
\newblock ~(2007).

\bibitem{gross_evenly_2007}
D.~Gross, K.~Audenaert, and J.~Eisert.
\newblock ``Evenly distributed unitaries: {On} the structure of unitary
  designs''.
\newblock \href{https://dx.doi.org/10.1063/1.2716992}{Journal of Mathematical
  Physics {\bf 48}, 052104}~(2007).

\bibitem{low_pseudo-randomness_2010}
Richard~A. Low.
\newblock ``Pseudo-randomness and {Learning} in {Quantum}
  {Computation}''~(2010).
\newblock  \href{http://arxiv.org/abs/1006.5227}{arXiv:1006.5227}.

\bibitem{akila_particle-time_2016}
M.~Akila, D.~Waltner, B.~Gutkin, and T.~Guhr.
\newblock ``Particle-time duality in the kicked {Ising} spin chain''.
\newblock \href{https://dx.doi.org/10.1088/1751-8113/49/37/375101}{Journal of
  Physics A: Mathematical and Theoretical {\bf 49}, 375101}~(2016).

\bibitem{bertini_exact_2018}
Bruno Bertini, Pavel Kos, and Toma{\v z} Prosen.
\newblock ``Exact {Spectral} {Form} {Factor} in a {Minimal} {Model} of
  {Many}-{Body} {Quantum} {Chaos}''.
\newblock \href{https://dx.doi.org/10.1103/PhysRevLett.121.264101}{Physical
  Review Letters {\bf 121}, 264101}~(2018).

\bibitem{bertini_exact_2019}
Bruno Bertini, Pavel Kos, and Toma\v{z} Prosen.
\newblock ``Exact {Correlation} {Functions} for {Dual}-{Unitary} {Lattice}
  {Models} in 1+1 {Dimensions}''.
\newblock \href{https://dx.doi.org/10.1103/PhysRevLett.123.210601}{Phys. Rev.
  Lett. {\bf 123}, 210601}~(2019).

\bibitem{gopalakrishnan_unitary_2019}
Sarang Gopalakrishnan and Austen Lamacraft.
\newblock ``Unitary circuits of finite depth and infinite width from quantum
  channels''.
\newblock \href{https://dx.doi.org/10.1103/PhysRevB.100.064309}{Physical Review
  B {\bf 100}, 064309}~(2019).

\bibitem{claeys_maximum_2020}
Pieter~W. Claeys and Austen Lamacraft.
\newblock ``Maximum velocity quantum circuits''.
\newblock \href{https://dx.doi.org/10.1103/PhysRevResearch.2.033032}{Physical
  Review Research {\bf 2}, 033032}~(2020).

\bibitem{claeys_emergent_2022}
Pieter~W. Claeys and Austen Lamacraft.
\newblock ``Emergent quantum state designs and biunitarity in dual-unitary
  circuit dynamics''.
\newblock \href{https://dx.doi.org/10.22331/q-2022-06-15-738}{Quantum {\bf 6},
  738}~(2022).

\bibitem{jozsa_lower_1994}
Richard Jozsa, Daniel Robb, and William~K. Wootters.
\newblock ``Lower bound for accessible information in quantum mechanics''.
\newblock \href{https://dx.doi.org/10.1103/PhysRevA.49.668}{Physical Review A
  {\bf 49}, 668--677}~(1994).

\bibitem{goldstein_distribution_2006}
Sheldon Goldstein, Joel~L. Lebowitz, Roderich Tumulka, and Nino Zangh{\`i}.
\newblock ``On the {Distribution} of the {Wave} {Function} for {Systems} in
  {Thermal} {Equilibrium}''.
\newblock \href{https://dx.doi.org/10.1007/s10955-006-9210-z}{Journal of
  Statistical Physics {\bf 125}, 1193--1221}~(2006).

\bibitem{goldstein_universal_2016}
Sheldon Goldstein, Joel~L. Lebowitz, Christian Mastrodonato, Roderich Tumulka,
  and Nino Zangh{\`i}.
\newblock ``Universal {Probability} {Distribution} for the {Wave} {Function} of
  a {Quantum} {System} {Entangled} with its {Environment}''.
\newblock \href{https://dx.doi.org/10.1007/s00220-015-2536-0}{Communications in
  Mathematical Physics {\bf 342}, 965--988}~(2016).

\bibitem{piroli_exact_2020}
Lorenzo Piroli, Bruno Bertini, J.~Ignacio Cirac, and Toma{\v z} Prosen.
\newblock ``Exact dynamics in dual-unitary quantum circuits''.
\newblock \href{https://dx.doi.org/10.1103/PhysRevB.101.094304}{Physical Review
  B {\bf 101}, 094304}~(2020).

\bibitem{anza_beyond_2021}
Fabio Anza and James~P. Crutchfield.
\newblock ``Beyond density matrices: {Geometric} quantum states''.
\newblock \href{https://dx.doi.org/10.1103/PhysRevA.103.062218}{Physical Review
  A {\bf 103}, 062218}~(2021).

\bibitem{anza_quantum_2022}
Fabio Anza and James~P. Crutchfield.
\newblock ``Quantum {Information} {Dimension} and {Geometric} {Entropy}''.
\newblock \href{https://dx.doi.org/10.1103/PRXQuantum.3.020355}{PRX Quantum
  {\bf 3}, 020355}~(2022).

\bibitem{renes_symmetric_2004}
Joseph~M. Renes, Robin Blume-Kohout, A.~J. Scott, and Carlton~M. Caves.
\newblock ``Symmetric informationally complete quantum measurements''.
\newblock \href{https://dx.doi.org/10.1063/1.1737053}{Journal of Mathematical
  Physics {\bf 45}, 2171--2180}~(2004).

\bibitem{knill_randomized_2008}
E.~Knill, D.~Leibfried, R.~Reichle, J.~Britton, R.~B. Blakestad, J.~D. Jost,
  C.~Langer, R.~Ozeri, S.~Seidelin, and D.~J. Wineland.
\newblock ``Randomized benchmarking of quantum gates''.
\newblock \href{https://dx.doi.org/10.1103/PhysRevA.77.012307}{Physical Review
  A {\bf 77}, 012307}~(2008).

\bibitem{roberts_chaos_2017}
Daniel~A. Roberts and Beni Yoshida.
\newblock ``Chaos and complexity by design''.
\newblock \href{https://dx.doi.org/10.1007/JHEP04(2017)121}{Journal of High
  Energy Physics {\bf 2017}, 121}~(2017).

\bibitem{banuls_matrix_2009}
M.~C. Ba{\~n}uls, M.~B. Hastings, F.~Verstraete, and J.~I. Cirac.
\newblock ``Matrix {Product} {States} for {Dynamical} {Simulation} of
  {Infinite} {Chains}''.
\newblock \href{https://dx.doi.org/10.1103/PhysRevLett.102.240603}{Phys. Rev.
  Lett. {\bf 102}, 240603}~(2009).

\bibitem{hastings_connecting_2015}
M.~B. Hastings and R.~Mahajan.
\newblock ``Connecting entanglement in time and space: {Improving} the folding
  algorithm''.
\newblock \href{https://dx.doi.org/10.1103/PhysRevA.91.032306}{Phys. Rev. A
  {\bf 91}, 032306}~(2015).

\bibitem{lerose_influence_2021}
Alessio Lerose, Michael Sonner, and Dmitry~A. Abanin.
\newblock ``Influence {Matrix} {Approach} to {Many}-{Body} {Floquet}
  {Dynamics}''.
\newblock \href{https://dx.doi.org/10.1103/PhysRevX.11.021040}{Physical Review
  X {\bf 11}, 021040}~(2021).

\bibitem{sonner_influence_2021}
Michael Sonner, Alessio Lerose, and Dmitry~A. Abanin.
\newblock ``Influence functional of many-body systems: {Temporal} entanglement
  and matrix-product state representation''.
\newblock \href{https://dx.doi.org/10.1016/j.aop.2021.168677}{Annals of Physics
  {\bf 435}, 168677}~(2021).

\bibitem{giudice_temporal_2022}
Giacomo Giudice, Giuliano Giudici, Michael Sonner, Julian Thoenniss, Alessio
  Lerose, Dmitry~A. Abanin, and Lorenzo Piroli.
\newblock ``Temporal {Entanglement}, {Quasiparticles}, and the {Role} of
  {Interactions}''.
\newblock \href{https://dx.doi.org/10.1103/PhysRevLett.128.220401}{Physical
  Review Letters {\bf 128}, 220401}~(2022).

\bibitem{ippoliti_postselection-free_2021}
Matteo Ippoliti and Vedika Khemani.
\newblock ``Postselection-{Free} {Entanglement} {Dynamics} via {Spacetime}
  {Duality}''.
\newblock \href{https://dx.doi.org/10.1103/PhysRevLett.126.060501}{Physical
  Review Letters {\bf 126}, 060501}~(2021).

\bibitem{ippoliti_fractal_2022}
Matteo Ippoliti, Tibor Rakovszky, and Vedika Khemani.
\newblock ``Fractal, {Logarithmic}, and {Volume}-{Law} {Entangled} {Nonthermal}
  {Steady} {States} via {Spacetime} {Duality}''.
\newblock \href{https://dx.doi.org/10.1103/PhysRevX.12.011045}{Physical Review
  X {\bf 12}, 011045}~(2022).

\bibitem{lu_spacetime_2021}
Tsung-Cheng Lu and Tarun Grover.
\newblock ``Spacetime duality between localization transitions and
  measurement-induced transitions''.
\newblock \href{https://dx.doi.org/10.1103/PRXQuantum.2.040319}{PRX Quantum
  {\bf 2}, 040319}~(2021).

\bibitem{stephen_universal_2022}
David~T. {Stephen}, Wen~Wei {Ho}, Tzu-Chieh {Wei}, Robert {Raussendorf}, and
  Ruben {Verresen}.
\newblock ``{Universal measurement-based quantum computation in a
  one-dimensional architecture enabled by dual-unitary circuits}''~(2022).
\newblock  \href{http://arxiv.org/abs/2209.06191}{arXiv:2209.06191}.

\bibitem{pollock_non-markovian_2018}
Felix~A. Pollock, Cesar Rodriguez-Rosario, Thomas Frauenheim, Mauro
  Paternostro, and Kavan Modi.
\newblock ``Non-{Markovian} quantum processes: {Complete} framework and
  efficient characterization''.
\newblock \href{https://dx.doi.org/10.1103/PhysRevA.97.012127}{Physical Review
  A {\bf 97}, 012127}~(2018).

\bibitem{kostenberger_weingarten_2021}
Georg K{\"o}stenberger.
\newblock ``Weingarten {Calculus}''~(2021).
\newblock  \href{http://arxiv.org/abs/2101.00921}{arXiv:2101.00921}.

\bibitem{kim_ballistic_2013}
Hyungwon Kim and David~A. Huse.
\newblock ``Ballistic {Spreading} of {Entanglement} in a {Diffusive}
  {Nonintegrable} {System}''.
\newblock \href{https://dx.doi.org/10.1103/PhysRevLett.111.127205}{Physical
  Review Letters {\bf 111}, 127205}~(2013).

\bibitem{nahum_quantum_2017}
Adam Nahum, Jonathan Ruhman, Sagar Vijay, and Jeongwan Haah.
\newblock ``Quantum {Entanglement} {Growth} under {Random} {Unitary}
  {Dynamics}''.
\newblock \href{https://dx.doi.org/10.1103/PhysRevX.7.031016}{Physical Review X
  {\bf 7}, 031016}~(2017).

\bibitem{jonay_coarse-grained_2018}
Cheryne Jonay, David~A. Huse, and Adam Nahum.
\newblock ``Coarse-grained dynamics of operator and state
  entanglement''~(2018).
\newblock  \href{http://arxiv.org/abs/1803.00089}{arXiv:1803.00089}.

\bibitem{zhou_emergent_2019}
Tianci Zhou and Adam Nahum.
\newblock ``Emergent statistical mechanics of entanglement in random unitary
  circuits''.
\newblock \href{https://dx.doi.org/10.1103/PhysRevB.99.174205}{Phys. Rev. B
  {\bf 99}, 174205}~(2019).

\bibitem{zhou_entanglement_2020}
Tianci Zhou and Adam Nahum.
\newblock ``Entanglement {Membrane} in {Chaotic} {Many}-{Body} {Systems}''.
\newblock \href{https://dx.doi.org/10.1103/PhysRevX.10.031066}{Phys. Rev. X
  {\bf 10}, 031066}~(2020).

\bibitem{li_statistical_2021}
Yaodong Li and Matthew P.~A. Fisher.
\newblock ``Statistical mechanics of quantum error correcting codes''.
\newblock \href{https://dx.doi.org/10.1103/PhysRevB.103.104306}{Physical Review
  B {\bf 103}, 104306}~(2021).

\bibitem{bensa_fastest_2021}
Jac Bensa and Marko Znidaric.
\newblock ``Fastest {Local} {Entanglement} {Scrambler}, {Multistage}
  {Thermalization}, and a {Non}-{Hermitian} {Phantom}''.
\newblock \href{https://dx.doi.org/10.1103/PhysRevX.11.031019}{Physical Review
  X {\bf 11}, 031019}~(2021).

\bibitem{bensa_two-step_2022}
Jac Bensa and Marko Znidaric.
\newblock ``Two-step phantom relaxation of out-of-time-ordered correlations in
  random circuits''.
\newblock \href{https://dx.doi.org/10.1103/PhysRevResearch.4.013228}{Physical
  Review Research {\bf 4}, 013228}~(2022).

\bibitem{wilming_high-temperature_2022}
Henrik Wilming and Ingo Roth.
\newblock ``High-temperature thermalization implies the emergence of quantum
  state designs''~(2022).
\newblock  \href{http://arxiv.org/abs/2202.01669}{arXiv:2202.01669}.

\bibitem{rakovszky_diffusive_2018}
Tibor Rakovszky, Frank Pollmann, and C.~W. von Keyserlingk.
\newblock ``Diffusive {Hydrodynamics} of {Out}-of-{Time}-{Ordered}
  {Correlators} with {Charge} {Conservation}''.
\newblock \href{https://dx.doi.org/10.1103/PhysRevX.8.031058}{Phys. Rev. X {\bf
  8}, 031058}~(2018).

\bibitem{khemani_operator_2018}
Vedika Khemani, Ashvin Vishwanath, and David~A. Huse.
\newblock ``Operator {Spreading} and the {Emergence} of {Dissipative}
  {Hydrodynamics} under {Unitary} {Evolution} with {Conservation} {Laws}''.
\newblock \href{https://dx.doi.org/10.1103/PhysRevX.8.031057}{Phys. Rev. X {\bf
  8}, 031057}~(2018).

\bibitem{hunter-jones_operator_2018}
Nicholas Hunter-Jones.
\newblock ``Operator growth in random quantum circuits with symmetry''~(2018).
\newblock  \href{http://arxiv.org/abs/1812.08219}{arXiv:1812.08219}.

\bibitem{agrawal_entanglement_2021}
Utkarsh Agrawal, Aidan Zabalo, Kun Chen, Justin~H. Wilson, Andrew~C. Potter,
  J.~H. Pixley, Sarang Gopalakrishnan, and Romain Vasseur.
\newblock ``Entanglement and charge-sharpening transitions in {U}(1) symmetric
  monitored quantum circuits''~(2021).
\newblock  \href{http://arxiv.org/abs/2107.10279}{arXiv:2107.10279}.

\bibitem{gopalakrishnan_operator_2018}
Sarang Gopalakrishnan.
\newblock ``Operator growth and eigenstate entanglement in an interacting
  integrable {Floquet} system''.
\newblock \href{https://dx.doi.org/10.1103/PhysRevB.98.060302}{Physical Review
  B {\bf 98}, 060302}~(2018).

\bibitem{klobas_exact_2021}
Katja Klobas, Bruno Bertini, and Lorenzo Piroli.
\newblock ``Exact {Thermalization} {Dynamics} in the ``{Rule} 54'' {Quantum}
  {Cellular} {Automaton}''.
\newblock \href{https://dx.doi.org/10.1103/PhysRevLett.126.160602}{Physical
  Review Letters {\bf 126}, 160602}~(2021).

\bibitem{buca_rule_2021}
Berislav Bu{\v c}a, Katja Klobas, and Toma{\v z} Prosen.
\newblock ``Rule 54: exactly solvable model of nonequilibrium statistical
  mechanics''.
\newblock \href{https://dx.doi.org/10.1088/1742-5468/ac096b}{Journal of
  Statistical Mechanics: Theory and Experiment {\bf 2021}, 074001}~(2021).

\bibitem{singh_fredkin_2022}
Hansveer Singh, Romain Vasseur, and Sarang Gopalakrishnan.
\newblock ``The {Fredkin} staircase: {An} integrable system with a
  finite-frequency {Drude} peak''~(2022).
\newblock  \href{http://arxiv.org/abs/2205.08542}{arXiv:2205.08542}.

\bibitem{lucas_generalized_2022}
Maxime Lucas, Lorenzo Piroli, Jacopo De~Nardis, and Andrea De~Luca.
\newblock ``Generalized {Deep} {Thermalization} for {Free} {Fermions}''~(2022).
\newblock  \href{http://arxiv.org/abs/2207.13628}{arXiv:2207.13628}.

\bibitem{von_keyserlingk_operator_2018}
C.~W. von Keyserlingk, Tibor Rakovszky, Frank Pollmann, and S.~L. Sondhi.
\newblock ``Operator {Hydrodynamics}, {OTOCs}, and {Entanglement} {Growth} in
  {Systems} without {Conservation} {Laws}''.
\newblock \href{https://dx.doi.org/10.1103/PhysRevX.8.021013}{Physical Review X
  {\bf 8}, 021013}~(2018).

\end{thebibliography}

\end{document}